\documentclass[11pt]{article}

\usepackage[a4paper,margin=1in]{geometry}
\usepackage{amsmath,amssymb,bm}
\usepackage{booktabs}
\usepackage{graphicx}
\usepackage{caption}
\usepackage{subcaption}
\usepackage{microtype}
\usepackage{xcolor}
\usepackage[numbers,sort&compress]{natbib}
\usepackage{hyperref}
\newcommand{\doi}[1]{doi:\,\href{https://doi.org/#1}{\nolinkurl{#1}}}
\usepackage{enumitem}
\usepackage{placeins}
\usepackage{float}
\usepackage{flafter}
\graphicspath{{./}{figures/}}
\hypersetup{colorlinks=true,linkcolor=blue!55!black,citecolor=blue!55!black,
  urlcolor=blue!60!black,pdftitle={Finite-Distance Imaging of Relativistic Equilibrium Tori around Genuine Kerr--Bertotti--Robinson Black Holes},
  pdfauthor={Hao-Peng Yan, Xiang-Qian Li, and Xiao-Jun Yue}}

\newcommand{\dd}{\mathrm{d}}

\title{Finite-Distance Imaging of Relativistic Equilibrium Tori around Genuine Kerr--Bertotti--Robinson Black Holes}
\author{Hao-Peng Yan$^{1,*}$, Xiang-Qian Li$^{1,*}$, and Xiao-Jun Yue$^1$\\[0.55em]
\small $^1$College of Physics and Optoelectronic Engineering, Taiyuan University of Technology,\\[-0.1em]
\small Taiyuan 030024, China\\[0.55em]
\small $^*$Co-corresponding authors: \href{mailto:yanhaopeng@tyut.edu.cn}{yanhaopeng@tyut.edu.cn};
\href{mailto:lixiangqian@tyut.edu.cn}{lixiangqian@tyut.edu.cn}\\[-0.1em]
\small E-mail: \href{mailto:yuexiaojun@tyut.edu.cn}{yuexiaojun@tyut.edu.cn}}
\date{23 October 2026}

\begin{document}

\maketitle

\begin{abstract}
We calculate finite-distance images of stationary, pressure-supported tori
around genuine, electrically neutral Kerr--Bertotti--Robinson black holes using general
relativistic ray tracing with gray emission and absorption. The emitting
volume is bounded by a constant-specific-angular-momentum equilibrium
free surface, and the observer is placed at a radius of $50M$.
Along the investigated external-field sequence, integrated brightness
and apparent size both increase strongly as the field modifies the
equilibrium source and photon propagation. Higher torus filling has
a different effect: in a representative sequence, brightness grows by
a factor of $5.9$ while the intensity-weighted image size increases by
only $10\%$. Increasing viewing inclination enhances lateral brightness
contrast. Stronger self-absorption can coexist with higher absolute
brightness because emission and absorption scale together in the adopted
gray prescription. At common pressure center and filling, increasing the
dimensionless spin from $0.5$ to $0.95$ dims the retrograde sequence by
a factor of about $7$ as its equilibrium contracts, whereas prograde
brightness changes by only $2\%$. These contrasting responses show how
the equilibrium boundary, fluid motion, and optical depth jointly shape
the brightness, scale, and asymmetry of finite-distance torus images.
\end{abstract}


\section{Introduction}

Black-hole images resolve radiation from matter whose distribution,
motion, and opacity shape the observed brightness. Event Horizon Telescope
images and polarization measurements make this connection central to
the interpretation of horizon-scale observations
\cite{EHTM87,EHTSgrA,EHTPolarization,EHTMagnetic}.
The vacuum capture boundary supplies a geometric reference, while direct
and strongly lensed emission depend on the source and viewing geometry
\cite{Gralla:2019xty,Perlick:2021aok,ChenReview,CunhaReview,
Johannsen,GrallaLensing,HouKN,ChaelInner}.
Fluid motion and radiative physics also affect intensity and polarization
asymmetries \cite{GatesBlueshift,GellesPolar,Palumbo}.
For a pressure-supported source of finite thickness, its free surface
determines the emitting volume, and foreground matter can attenuate
radiation from deeper layers. Relating this equilibrium structure to the
image therefore requires volume radiative transfer as well as null-ray
propagation.

An external electromagnetic background adds a scale to the black-hole
environment, as illustrated by the test-field and exact magnetic-universe
constructions \cite{Wald,Ernst,ErnstWild}. External-field effects on optical
appearance have been studied through lensing, shadow, and emission
calculations \cite{Stuchlik,WangMelvin,JuniorMag,HouKerrMelvin,ZhuGuo,HuQED,ZhongQED}.
The Kerr--Bertotti--Robinson (KBR) solution describes a rotating black hole
in a Bertotti--Robinson electromagnetic background \cite{OldSolution,AlBadawi}.
We consider the electrically neutral solution, called the genuine KBR
solution in the literature, which is distinct from the earlier special-charge
configuration \cite{Genuine,Thermo,MaLuSeeds}.
Hereafter, KBR denotes this neutral solution. The geometry and thermodynamics
of the wider Kerr--Newman--BR family have been examined in
Refs.~\cite{GlobalKBR,ExtendedKNBR,KNBRThermo,NotesKBR}.
It recovers Kerr at zero external field but is not asymptotically flat at
nonzero field. In this work, we calculate images for a finite-distance
observer and impose a radiative boundary condition at the outer edge of
the ray-tracing domain. In this setting, varying the
field changes both photon propagation and the potential supporting the
emitting fluid.

Existing KBR and Kerr--Newman--BR optical calculations describe vacuum
capture boundaries, photon rings, and images of thin emitting sources
\cite{ShadowOld,OpticalOld,PhotonRing,ThinKNBR,Synchrotron}.
They establish propagation and surface-emission signatures, but do not
determine attenuation through a three-dimensional pressure-supported torus.
Equilibrium fluid models in other backgrounds
\cite{Cassing,QMetricTorus,KerrMOG} and thick-source imaging calculations
\cite{StraubIonTorus,VincentTorus,ZhangThick} provide a basis for treating
finite thickness. The neutral KBR geometry poses the coupled question of
how its equilibrium free surface, fluid motion, and volume absorption
affect the brightness and extent of the image received by a finite observer.

We address this question with a stationary, neutral, polytropic torus of
constant specific angular momentum, following the relativistic
equilibrium construction \cite{Fishbone,KozlowskiTorus,PolishTorus,
AbramowiczFragile,PennaTorus}. The BR field enters through the metric;
the source has no internal magnetic pressure. Gray emission and absorption
are integrated through the equilibrium volume using invariant radiative
transfer \cite{Lindquist,Younsi,DavisGammie}. The comparisons determine how
brightness and image size respond to field strength and filling, how
inclination and optical depth redistribute the received radiation, and
how spin and rotation sense alter the source structure and its image.
The finite-distance construction fixes the observer and radiative boundary
for each comparison rather than using asymptotic screen coordinates.

Section~II specifies the spacetime and ray tracing, Sec.~III gives the
source and transfer model, and Sec.~IV presents the images and their
brightness, size, and asymmetry. Section~V summarizes the physical results and their limitations.

\section{KBR spacetime and ray tracing}
\subsection{KBR geometry and null geodesics}

In Boyer--Lindquist-type coordinates $(t,r,\theta,\phi)$, the electrically neutral
Kerr--Bertotti--Robinson metric takes the form \cite{Genuine,Thermo}
\begin{align}
 \dd s^2=\Omega^{-2}\bigg[&-\frac{Q}{\Sigma}
   (\dd t-a\sin^2\theta\,\dd\phi)^2
   +\frac{\Sigma}{Q}\,\dd r^2
   +\frac{\Sigma}{P}\,\dd\theta^2 \\[-2pt]
   &+\frac{P\sin^2\theta}{\Sigma}
   \left(a\,\dd t-A\,\dd\phi\right)^2\bigg],
\label{eq:metric}
\end{align}
where the metric functions are
\begin{align}
 \Sigma &= r^2+a^2\cos^2\theta, &
 \Delta &= r^2-2Mr+a^2,\\
 P &= 1-\tilde B^2(M^2-a^2)\sin^2\theta, &
 Q &= \Delta\left(1+\tilde B^2\Delta\right),\\
 \Omega^2 &= 1+\tilde B^2\Delta\sin^2\theta, &
 A &= r^2+a^2 .
\label{eq:metric-functions}
\end{align}
Here $M$ is the mass parameter, $a$ is the rotation parameter, and
$\tilde B$ characterizes the external Bertotti--Robinson electromagnetic
field. We use geometrized units $G=c=1$; $M$ and $a$ have dimensions of
length, while $\tilde B$ has dimensions of inverse length. The coordinates
and parameters follow the physical parametrization of Ref.~\cite{Thermo},
with coordinate tildes omitted. The regular-axis azimuth has period $2\pi$.
We impose $|a|\leq M$ and $\tilde B^2(M^2-a^2)<1$, the latter ensuring
$P>0$ for all polar angles. The outer horizon is at
\begin{equation}
 r_+=M+\sqrt{M^2-a^2}.
\end{equation}
The familiar Kerr limit is recovered at $\tilde B=0$.  At nonzero $\tilde B$ the geometry has BR asymptotics, and the optical construction therefore specifies both observer and source at finite radius.

The nonzero covariant metric components are
\begin{align}
 g_{tt}&=\frac{-Q+Pa^2\sin^2\theta}{\Omega^2\Sigma},\\
 g_{t\phi}&=\frac{a\sin^2\theta(Q-PA)}{\Omega^2\Sigma},\\
 g_{\phi\phi}&=\frac{\sin^2\theta(PA^2-Qa^2\sin^2\theta)}{\Omega^2\Sigma},\\
 g_{rr}&=\frac{\Sigma}{\Omega^2Q},\qquad
 g_{\theta\theta}=\frac{\Sigma}{\Omega^2P}.
\end{align}
The metric components satisfy
\begin{equation}
 D\equiv g_{t\phi}^2-g_{tt}g_{\phi\phi}
   =\frac{QP\sin^2\theta}{\Omega^4}.
\label{eq:detidentity}
\end{equation}

The Hamilton--Jacobi action for a null ray separates, following the Carter
construction and the quadratic-integral description of null motion
\cite{Carter,WalkerPenrose}, applied to this neutral KBR branch
as in Ref.~\cite{Thermo}.
Let $E=-p_t$ and $L=p_\phi$ be the conserved Killing energy and axial angular
momentum, and let $\mathcal K$ be the separation constant. We introduce
the energy-normalized constants
\begin{equation}
 \xi=\frac{L}{E},\qquad
 \eta=\frac{\mathcal K}{E^2}-(\xi-a)^2.
\label{eq:ray-constants}
\end{equation}
In the Kerr limit, $E^2\eta$ is the usual Carter constant. The first-order
equations in the Mino-like parameter $\gamma$ are
\begin{align}
 \frac{\dd r}{\dd\gamma}&=\pm\sqrt{R(r)}, &
 \frac{\dd\theta}{\dd\gamma}&=\pm\sqrt{\Theta(\theta)},\\
 \frac{\dd t}{\dd\gamma}&=\frac{A(AE-aL)}{Q}
 +\frac{a(L-aE\sin^2\theta)}{P},\\
 \frac{\dd\phi}{\dd\gamma}&=\frac{a(AE-aL)}{Q}
 +\frac{L-aE\sin^2\theta}{P\sin^2\theta}.
\label{eq:null-geodesics}
\end{align}
The radial and polar potentials appearing in these equations are
\begin{align}
 R(r)&=(AE-aL)^2-QE^2[\eta+(\xi-a)^2],\\
 \Theta(\theta)&=PE^2[\eta+(\xi-a)^2]
 -\frac{(L-aE\sin^2\theta)^2}{\sin^2\theta}.
\label{eq:potentials}
\end{align}
The conformal factor cancels from the unparameterized null path but remains in
its affine normalization: $\dd\lambda=\Sigma\,\dd\gamma/\Omega^2$, where
$\lambda$ is an affine parameter. The variable $\mu=\cos\theta$ gives a
regular form at the polar axis. After normalizing $E=1$, with
$C=\tilde B^2(M^2-a^2)$, the polar potential and regularized equations are
\begin{equation}
\begin{aligned}
 U(\mu)&=(1-\mu^2)[1-C(1-\mu^2)][\eta+(\xi-a)^2]
          -[\xi-a(1-\mu^2)]^2,\\
 \mu'^2&=U,\qquad \mu''=\tfrac12 U_{,\mu},\qquad r''=\tfrac12 R_{,r}.
\end{aligned}
\end{equation}
Here primes on the trajectory variables denote differentiation with respect
to $\gamma$, and a comma denotes a coordinate derivative.
This polynomial system crosses polar turning points smoothly because it contains no division by $\sin\theta$.

\subsection{Observer's sky and backward ray tracing}

We place a zero-angular-momentum observer at $(r_o,\theta_o)$ with the baseline
radius $r_o=50M$ and inclination $i\equiv\theta_o$, measured from the rotation
axis. The most extended equilibrium in the pressure-center survey
has $r_{\rm out}=27.592M$, so this choice places the observer outside every
emitting volume. Appendix~\ref{app:observer-radius} quantifies the remaining
dependence on $r_o$. The local observer and orthonormal spatial triad follow
the standard finite-distance construction \cite{Bardeen,Perlick:2021aok,ChangObserver}:
\begin{align}
 u_o&=e_{(t)}=\frac{1}{N_o}(\partial_t+\omega_o\partial_\phi),
 &e_{(r)}&=\frac{\partial_r}{\sqrt{g_{rr}}},\nonumber\\
 e_{(\theta)}&=\frac{\partial_\theta}{\sqrt{g_{\theta\theta}}},
 &e_{(\phi)}&=\frac{\partial_\phi}{\sqrt{g_{\phi\phi}}},
\label{eq:zamo}
\end{align}
where $N_o^2=-g_{tt}+g_{t\phi}^2/g_{\phi\phi}$ and $\omega_o=-g_{t\phi}/g_{\phi\phi}$ are evaluated at the observer. This local frame is required because the BR exterior does not provide an asymptotically inertial screen.

Let $(x,y)$ be gnomonic screen coordinates and ${\cal N}=\sqrt{1+x^2+y^2}$. We plot $\alpha=r_ox$ and $\beta=r_oy$, with positive $\beta$ pointing toward decreasing Boyer--Lindquist polar angle. The backward spatial tracing direction has tetrad components
\begin{equation}
 n^{(r)}=-\frac{1}{\cal N},\qquad
 n^{(\theta)}=-\frac{y}{\cal N},\qquad
 n^{(\phi)}=-\frac{x}{\cal N}.
\label{eq:screen-direction}
\end{equation}
For unit local frequency before the Killing-energy rescaling, define
\begin{equation}
 L_0=\sqrt{g_{\phi\phi,o}}\,n^{(\phi)},\qquad
 E_0=N_o+\omega_oL_0.
\end{equation}
The constants used by the separated equations are then normalized to $E=1$:
\begin{align}
 \xi&=\frac{L_0}{E_0},\qquad
 \frac{p_\theta}{E}=\frac{\sqrt{g_{\theta\theta,o}}\,n^{(\theta)}}{E_0},\nonumber\\
 \eta&=P_o\left(\frac{p_\theta}{E}\right)^2
 +\frac{(\xi-a\sin^2\theta_o)^2}{P_o\sin^2\theta_o}-(\xi-a)^2.
\label{eq:screen-map}
\end{align}
Equations~\eqref{eq:screen-direction} and \eqref{eq:screen-map} map each
screen pixel to $(\xi,\eta)$. The initial Mino derivatives follow from the
same tetrad projection,
\begin{equation}
 r'_o=-\frac{\Sigma_o}{\Omega_o^2\sqrt{g_{rr,o}}\,{\cal N}E_0},\qquad
 \theta'_o=-\frac{\Sigma_o y}{\Omega_o^2\sqrt{g_{\theta\theta,o}}\,{\cal N}E_0}.
\label{eq:initial-derivatives}
\end{equation}
Thus every image pixel defines one backward ray without an intermediate asymptotic-impact-parameter approximation.

For spherical photon orbits, solving $R=R_{,r}=0$ for the potentials in
Eq.~\eqref{eq:potentials} gives, at $a\ne0$,
\begin{equation}
 \xi_{\rm sph}=\frac{r^2+a^2-4rQ/Q_{,r}}{a},\qquad
 \eta_{\rm sph}=\frac{16r^2Q}{(Q_{,r})^2}-(\xi_{\rm sph}-a)^2.
\end{equation}
We retain unstable orbits that are radially accessible from the observer and project their constants through Eq.~\eqref{eq:screen-map}. The resulting vacuum capture boundaries appear as dashed curves in the intensity images. Backward rays terminate at the horizon or at their first outward recrossing of $r_o$; the latter carries zero incident intensity. These conditions specify the transfer boundary values on the finite imaging domain.

\section{Equilibrium torus and radiative transfer}
\subsection{Circular flow and the effective potential}

We model a stationary, axisymmetric, circular neutral test fluid,
$u^\mu=u^t(1,0,0,\Omega_f)$.  Its specific angular momentum and angular
velocity are
\begin{equation}
 \ell=-\frac{u_\phi}{u_t},\qquad \Omega_f=\frac{u^\phi}{u^t}.
\end{equation}
For a barotropic fluid the meridional Euler equation has the standard form
\cite{Fishbone,KozlowskiTorus,PolishTorus,AbramowiczFragile}
\begin{equation}
 \frac{\partial_i p}{e+p}=-\partial_i\ln|u_t|
 +\frac{\Omega_f\partial_i\ell}{1-\Omega_f\ell},\qquad i\in\{r,\theta\}.
\label{eq:euler}
\end{equation}
Defining
\begin{equation}
 {\cal A}=g_{\phi\phi}+2\ell g_{t\phi}+\ell^2g_{tt},
 \qquad D=g_{t\phi}^2-g_{tt}g_{\phi\phi},
\end{equation}
the constant-$\ell$ solution has the standard Polish-doughnut potential
\cite{Fishbone,KozlowskiTorus,PolishTorus,TlemissovKovar}
\begin{equation}
 W\equiv\ln|u_t|=\frac12\ln\left|\frac{D}{\mathcal A}\right|.
\label{eq:potential}
\end{equation}
The circular four-velocity is therefore fixed algebraically,
\begin{align}
 \Omega_f&=-\frac{g_{t\phi}+\ell g_{tt}}
 {g_{\phi\phi}+\ell g_{t\phi}},\\
 u^t&=\sqrt{\frac{D}{\mathcal A}}
 \frac{g_{\phi\phi}+\ell g_{t\phi}}{D}.
\label{eq:fluid}
\end{align}
The physical torus is restricted to the connected region containing the
chosen pressure center where the circular velocity is timelike and $D>0$,
$\mathcal A>0$.

We prescribe an equatorial center $r_c$ and set $\ell$ equal to the specific
angular momentum of the corresponding circular geodesic. Explicitly, the
standard circular-orbit relations give \cite{Fishbone,PennaTorus,WangISCO}
\begin{align}
 \Omega_{\rm K}^{\pm}&=
 \frac{-g_{t\phi,r}\pm\sqrt{g_{t\phi,r}^2-g_{tt,r}g_{\phi\phi,r}}}
 {g_{\phi\phi,r}},\\
 \ell_{\rm K}^{\pm}&=-\frac{g_{t\phi}+g_{\phi\phi}\Omega_{\rm K}^{\pm}}
 {g_{tt}+g_{t\phi}\Omega_{\rm K}^{\pm}},\qquad
 \ell=\ell_{\rm K}^{\pm}(r_c,\pi/2).
\label{eq:circular-center}
\end{align}
The subscript $\mathrm K$ denotes the Keplerian (circular-geodesic)
angular velocity or specific angular momentum. All metric derivatives
in these relations are evaluated on the equator;
for the positive spins used here, the plus and minus branches correspond to
prograde and retrograde motion. The pressure center is a local minimum of
$W$ on the equatorial plane, with potential value $W_c\equiv W(r_c,\pi/2)$.
The neighboring saddle of $W$ defines a cusp. We denote by $W_{\rm crit}$
the lowest potential level at which the connected lobe around the center
opens; for the fiducial sequence, this is the potential at the inner cusp.
For a closed equilibrium without cusp overflow, we choose a surface
potential $W_s$ between $W_c$ and $W_{\rm crit}$:
\begin{equation}
 W_s=W_c+\chi(W_{\rm crit}-W_c),\qquad 0<\chi<1,
\label{eq:filling}
\end{equation}
The zero-pressure free surface is the connected equipotential $W(r,\theta)=W_s$
enclosing $r_c$; the fluid occupies its interior, where $W<W_s$.
The filling parameter $\chi=(W_s-W_c)/(W_{\rm crit}-W_c)$ specifies the
fraction of the available potential depth occupied by the torus. As
$\chi$ increases from zero toward unity, the closed surface expands from
the pressure center toward the critical opening surface.

\begin{figure}[!htbp]
\centering
\includegraphics[width=0.98\linewidth]{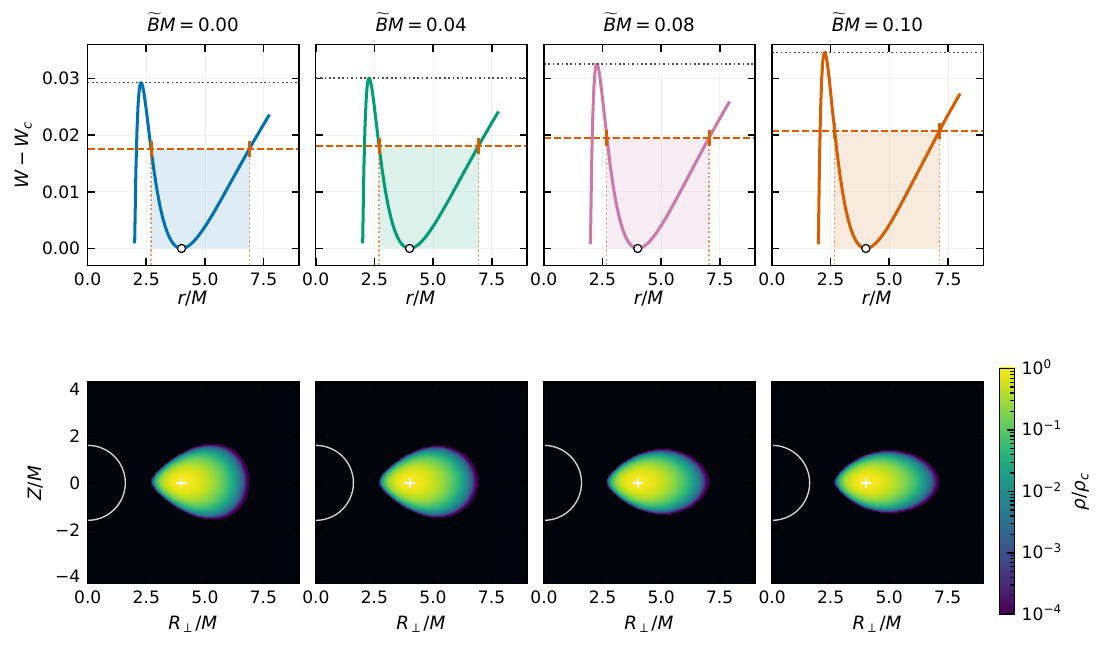}
\caption{Construction of the fiducial equilibria at $a/M=0.8$, $r_c=4M$,
and $\chi=0.60$.  Columns use the same field sequence as the image survey.
The upper row shows $W-W_c$, the selected surface (dashed), and the critical
barrier (dotted); vertical marks are the equatorial surface intersections.
The lower row shows the density obtained from the same potential.  The light
semicircle is the coordinate horizon and the cross marks $r_c$.  On the
equatorial plane $Z=0$, $R_\perp=r$, so the upper-row intersections coincide with
the inner and outer equatorial edges of the corresponding lower-row torus.}
\label{fig:potentialconstruction}
\end{figure}

For the spatial plots, we use the auxiliary coordinates
$X=r\sin\theta\cos\phi$, $Y=r\sin\theta\sin\phi$, and $Z=r\cos\theta$.
The meridional density sections are plotted in
$(R_\perp,Z)$, where $R_\perp=\sqrt{X^2+Y^2}=r\sin\theta$.
These are spatial plotting coordinates, not proper distances or the
observer's screen coordinates $(x,y)$. On the equatorial plane $Z=0$,
$R_\perp=r$; the surface intersections in the upper panels of
Fig.~\ref{fig:potentialconstruction} therefore match the equatorial torus
edges in the lower panels, which use the same horizontal range.

At fixed $r_c$ and $\chi$, the external field changes the potential depth
and both equatorial surface intersections
(Fig.~\ref{fig:potentialconstruction}). The resulting density profiles and
photon propagation therefore both vary along the field sequence.

\subsection{Polytropic closure and local radiative weights}

We adopt $p=K_{\rm p}\rho^\Gamma$ and $e=\rho+p/(\Gamma-1)$, where
$\rho$ is rest-mass density, $e$ is total rest-frame energy density, and
$K_{\rm p}$ is the polytropic constant.
The specific enthalpy is \cite{PolishTorus,PennaTorus}
\begin{equation}
 h=\frac{e+p}{\rho}=1+\frac{\Gamma}{\Gamma-1}K_{\rm p}\rho^{\Gamma-1}.
\end{equation}
Integrating Eq.~\eqref{eq:euler} for constant $\ell$ and imposing $p=0$ on
$W=W_s$ gives
\begin{align}
 h(r,\theta)&=\exp(W_s-W),\\
 \frac{\rho}{\rho_c}&=\left(\frac{h-1}{h_c-1}\right)^{1/(\Gamma-1)},
 \qquad h_c=\exp(W_s-W_c),
\label{eq:polytrope}
\end{align}
inside the connected lobe, with $h=1$ and $\rho=0$ outside.  For the
normalization $\rho_c=1$, the polytropic constant is
$K_{\rm p}=(\Gamma-1)(h_c-1)/\Gamma$. We set $\Gamma=4/3$ throughout.

\begin{figure}[!htbp]
\centering
\includegraphics[width=0.90\linewidth]{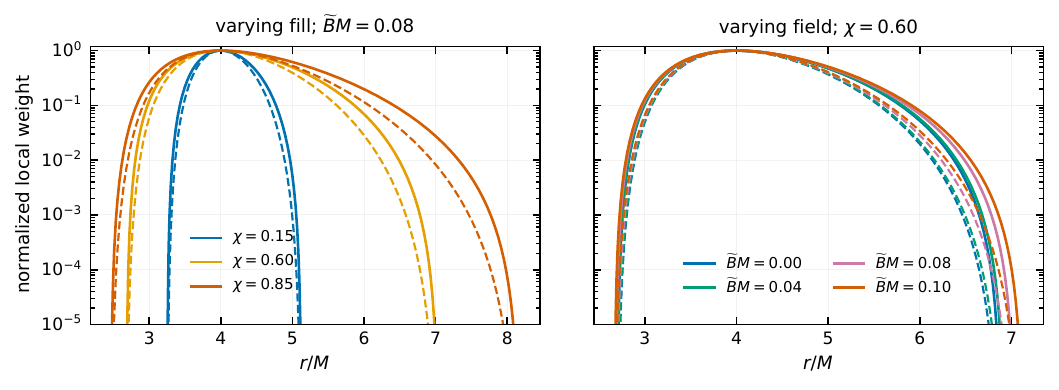}
\caption{Equatorial profiles that enter the transfer calculation.  Solid
curves show $\rho/\rho_c\propto\alpha_e$ and dashed curves show
$q^4=S_e/S_c$.  The left panel varies the filling level at fixed field; the
right varies the external field at fixed filling.  Every curve terminates at
the two free-surface intersections determined by Eq.~\eqref{eq:filling}.}
\label{fig:sourceprofiles}
\end{figure}

For the dimensionless gray coefficients, we define
\begin{equation}
 q\equiv\frac{h-1}{h_c-1},\qquad
 \frac{\rho}{\rho_c}=q^{1/(\Gamma-1)},
\end{equation}
we take $\alpha_e M=\tau_0\rho/\rho_c$ and $S_e/S_c=q^4$. Here $\alpha_e$
is the rest-frame absorption coefficient, $S_e=j_e/\alpha_e$ is the gray
source function, $j_e$ is the emission coefficient, and $S_c$ is the common
central source-function normalization. The opacity therefore tracks the equilibrium density, while the source
function is more centrally concentrated.  Figure~\ref{fig:sourceprofiles} displays both weights on the
equator and shows directly how changing $\chi$ or $\widetilde BM$ changes the
emitting interval.  At $\chi=0.60$, the solid curves in its right panel are
the equatorial cuts of the lower-row density maps in
Fig.~\ref{fig:potentialconstruction}; the dashed curves are the source-function
weights constructed from the same equilibrium variable $q$.  No independent
radial emissivity profile is imposed.

The torus is neutral and non-self-gravitating.  The BR field therefore enters
the equilibrium through the metric, not through a Lorentz force or internal
magnetic pressure.

\subsection{Frequency shift and gray transfer}

The observer and emitter measure the photon frequency through
$\nu=-p_\mu u^\mu$ (with a common frequency-unit convention), yielding
\cite{Lindquist,Younsi,DavisGammie}
\begin{align}
 \nu_o&=-p_\mu u_o^\mu=\frac{E(1-\omega_o\xi)}{N_o},
 \label{eq:observer-frequency}\\
 \nu_e&=-p_\mu u_e^\mu=Eu^t(1-\Omega_f\xi),\\
 g&\equiv\frac{\nu_o}{\nu_e}
 =\frac{(1-\omega_o\xi)/N_o}{u^t(1-\Omega_f\xi)}.
\label{eq:redshift}
\end{align}
Equation~\eqref{eq:observer-frequency} also explains the screen normalization
$E_0$: after setting $E=1$, $\nu_o=1/E_0$.  Along positive backward Mino
distance the proper path element is
$\dd\ell_{\rm prop}=\nu_e\Sigma\,\dd\gamma/\Omega^2$.  The formal gray
bolometric solution with zero incident intensity at the outer boundary is
\cite{Lindquist,Younsi,DavisGammie}
\begin{equation}
 \frac{I_o}{S_c}=\int g^4q^4 e^{-\tau}\,\dd\tau,
 \qquad \dd\tau=\alpha_e\,\dd\ell_{\rm prop}.
\label{eq:transfer}
\end{equation}
Here $\tau$ is measured from the observer to the emitting segment along
the backward ray. We denote the total optical depth through all traversed
fluid by $\tau_{\rm los}=\int\alpha_e\,\dd\ell_{\rm prop}$; it is distinct
from the coefficient normalization $\tau_0$.
The factor $g^4$ is the bolometric counterpart of the familiar $g^3$
specific-intensity transformation.  The common $S_c$ fixes the relative
normalization across models, but it does not provide a calibrated spectrum or
polarization prediction.

Each numerical segment contributes
$g^4q^4(1-e^{-\Delta\tau})e^{-\tau_{\rm foreground}}$, using midpoint fluid
variables and the exact cell absorption factor.  We also accumulate
$\int g\,\dd I_o$ and report the contribution-weighted frequency shift
$\langle g\rangle_I=\int g\,\dd I_o/\int\dd I_o$.

The dimensionless integrated brightness and intensity-weighted image moments
are defined on the observer's local sky by
\begin{align}
 \mathcal F&=\int (I_o/S_c)\,\dd\Omega_o,
 &\dd\Omega_o&=\frac{\dd x\,\dd y}{(1+x^2+y^2)^{3/2}},\\
 \bar\alpha&=\mathcal F^{-1}\int\alpha(I_o/S_c)\,\dd\Omega_o,
 &\sigma_\alpha^2&=\mathcal F^{-1}\int(\alpha-\bar\alpha)^2(I_o/S_c)\,\dd\Omega_o.
\label{eq:image-observables}
\end{align}
The corresponding $\beta$ moments are defined in the same way. The lateral
brightness contrast is measured by $\mathcal F_{\rm L}/\mathcal F_{\rm R}$,
where the two integrals cover $\alpha<0$ and $\alpha>0$, respectively, with
the same solid-angle measure.
$\mathcal F_{\rm thin}$ uses the same emissivity and rays with the foreground
attenuation factor removed. Thus $\mathcal F$ is a dimensionless local-sky
integral, and the plotted rms size $\sqrt{\sigma_\alpha^2+\sigma_\beta^2}$ is
conditional on the screen radius used to define $\alpha$ and $\beta$.

\subsection{Numerical protocol and comparison grid}

The regularized $(r,\mu)$ equations are advanced with fourth-order
Runge--Kutta steps limited by the local radial rate and
$\mathcal K/E^2=\eta+(\xi-a)^2$. Every ray records
its terminal state, torus intersection, accumulated optical depth, and the
maximum normalized violations of $r'^2=R$ and $\mu'^2=U$.  The main images use
$320^2$ pixels and the common $100^\circ$ field of view listed in
Table~\ref{tab:grid}; a separate common-center sequence is used only for the
rotation-sense comparison, with $384^2$ pixels at $a/M=0.5,0.8$ and
$768^2$ at $a/M=0.95$.

\begin{table}[!htbp]
\centering
\caption{Parameters of the fiducial comparison grid.}
\label{tab:grid}
\begin{tabular}{llp{6.0cm}}
\toprule
Quantity & Fiducial value & Values varied in figures\\
\midrule
$a/M$ & $0.8$ & fixed\\
$r_c/M$ & $4$ & fixed\\
$\Gamma$ & $4/3$ & fixed\\
$\widetilde BM$ & $0.08$ & $0,0.04,0.08,0.10$\\
$\chi$ & $0.60$ & $0.15,0.60,0.85$\\
$i$ & $60^\circ$ & $20^\circ,40^\circ,60^\circ,80^\circ$\\
$\tau_0$ & $0.20$ & $0.02,0.20,2.00$\\
$r_o/M$ & $50$ & $30,40,50,60,70,80$ in Appendix~\ref{app:observer-radius}\\
Image grid & $320^2$ & displayed matrices\\
Scalar-scan grid & $224^2$ & $448^2$ for the Kerr field scan\\
\bottomrule
\end{tabular}
\end{table}

The three displayed filling levels sample low, intermediate, and nearly
critical equilibria. The scalar scans resolve the trends with the denser samples
\begin{align*}
\chi&=(0.15,0.25,0.35,0.45,0.60,0.70,0.78,0.85),\\
i&=(20^\circ,30^\circ,40^\circ,50^\circ,60^\circ,70^\circ,80^\circ),\\
\tau_0&=(0.02,0.05,0.10,0.20,0.50,1.00,2.00).
\end{align*}
They are evaluated at $224^2$ pixels, with $448^2$ for the angularly compact
Kerr control in the field scan. Every point is obtained from
an independent ray-tracing calculation without interpolation.

For the fiducial $i=80^\circ$ model, halving the maximum integration step
leaves the $96^2$ terminal and torus-hit labels unchanged and changes
$\mathcal F$ by $2.2\times10^{-7}$ in relative terms. For the high-spin
common-center models, increasing the image grid from $384^2$ to $768^2$
changes $\mathcal F$ by $0.03\%$ (prograde) and $0.18\%$ (retrograde),
and rms size by less than $0.04\%$. The retrograde horizontal and vertical
cut maxima change by $1.3\%$ and $5.2\%$, respectively. These maxima are
more resolution sensitive than the integrated intensity and rms size, with residual
percent-level resolution dependence in the finite-pixel profile amplitudes.

\section{Numerical results and image analysis}
The images probe three related aspects of the source--propagation problem:
the response to external field and filling, the viewing and transfer
effects for a fixed equilibrium, and the coupled structural and kinematic
changes associated with spin and rotation sense. The field--filling and
inclination/opacity sequences use $a/M=0.8$, $r_c=4M$, and $r_o=50M$.
The pressure-center and rotation-sense surveys use separate comparison
families specified in Sec.~\ref{sec:spin-equilibrium}; their absolute
brightnesses should not be compared across families as single-parameter
variations. Within each image matrix, common axes and color normalization
permit direct comparison.

\subsection{External-field strength and torus filling}

\begin{figure}[!htbp]
\centering
\includegraphics[width=0.90\linewidth]{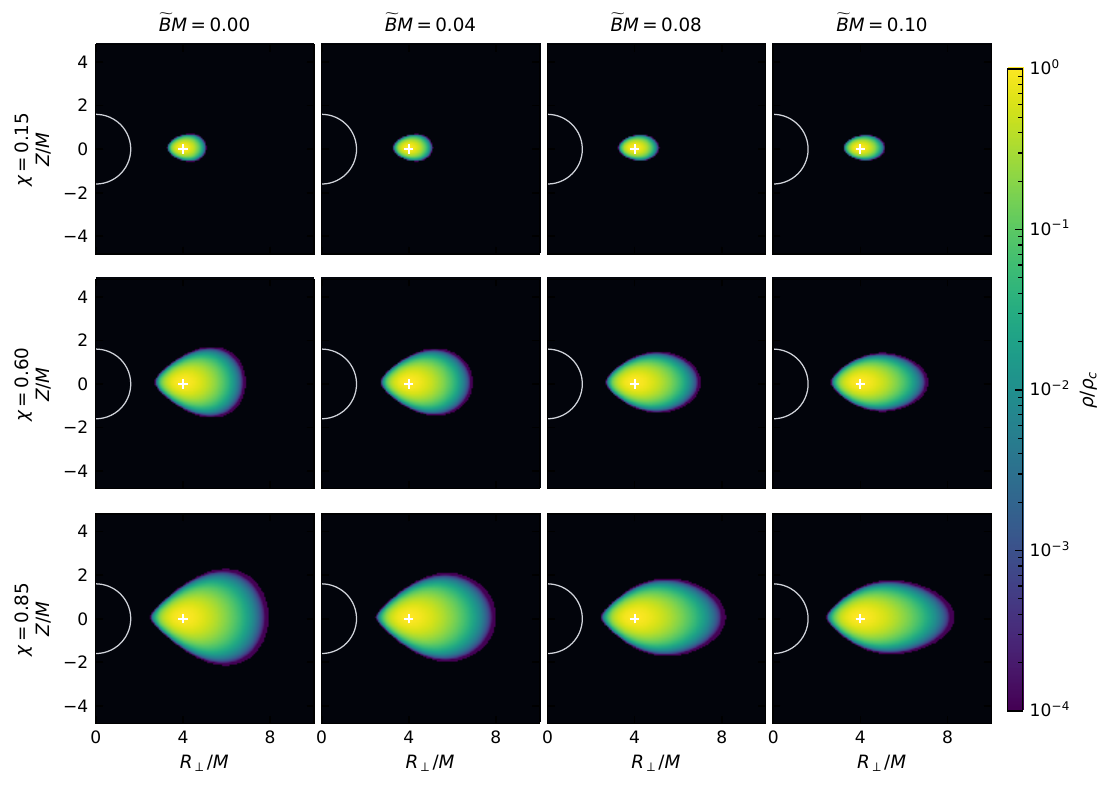}
\caption{Normalized meridional density for the matched equilibrium grid.
Rows give $\chi=(0.15,0.60,0.85)$ and columns give
$\widetilde BM=(0,0.04,0.08,0.10)$; $a/M=0.8$, $r_c=4M$, and
$\Gamma=4/3$.  All panels share the same coordinate window and logarithmic
color scale.}
\label{fig:torusmatrix}
\end{figure}

Figure~\ref{fig:torusmatrix} shows the equilibrium density across the
field--filling grid.  Increasing $\chi$ widens and
thickens the connected lobe.  Increasing the field at fixed $r_c$ and $\chi$
also changes the outer low-density envelope because the potential itself has
changed. The density vanishes outside the closed free surface, including
the gap between the horizon and the inner torus edge.

\begin{figure}[!htbp]
\centering
\includegraphics[width=0.90\linewidth]{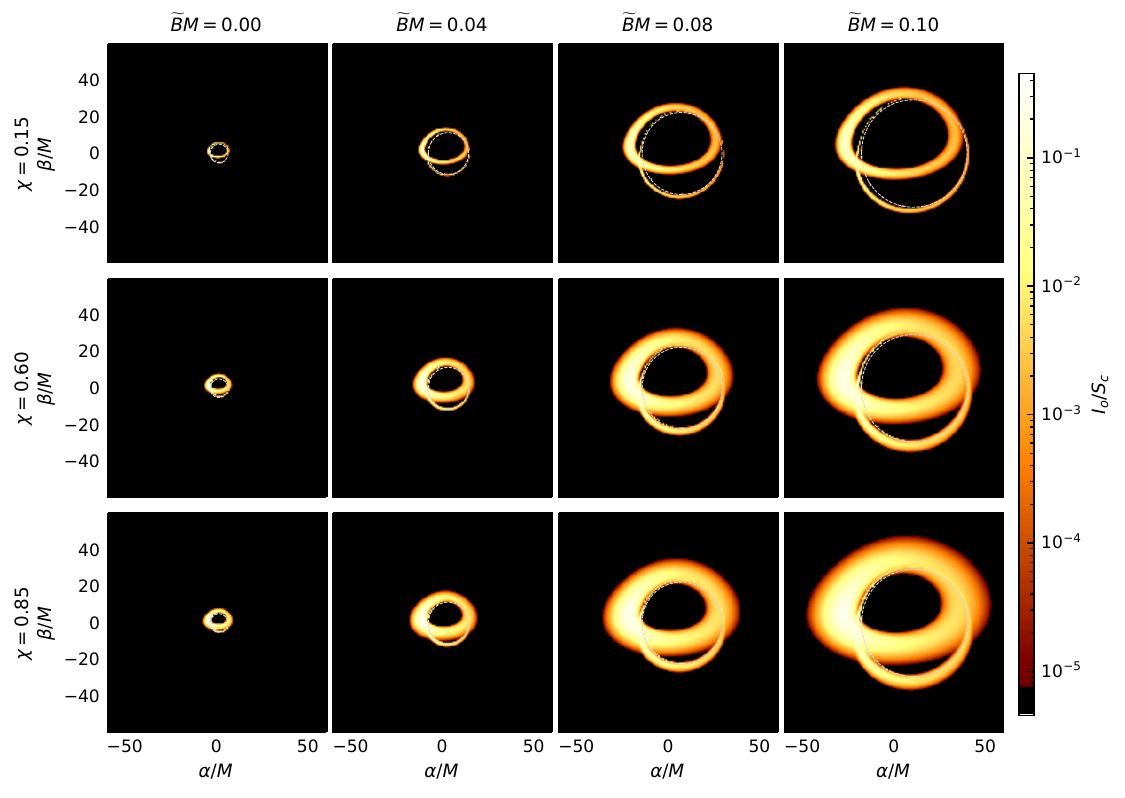}
\caption{Observed gray intensity for the same $(\chi,\widetilde BM)$ grid as
Fig.~\ref{fig:torusmatrix}, at $r_o=50M$, $i=60^\circ$, and $\tau_0=0.20$.  A single
absolute logarithmic scale is used throughout.  Dashed curves are the vacuum
critical curves and are references, not emitting surfaces.}
\label{fig:intensity-chib}
\end{figure}

The corresponding intensity sequence is shown in
Fig.~\ref{fig:intensity-chib}.  Its rows and columns correspond to the
equilibria in Fig.~\ref{fig:torusmatrix}.  Filling mainly changes the source
extent, whereas the BR field changes both that extent and the finite-distance
lens map.  The increasing image scale with field is therefore not a pure
shadow-size effect.

The prescribed circular-ring and celestial-source maps in
Appendix~\ref{app:vacuum} show the field dependence of photon propagation
without torus absorption.

\FloatBarrier

\subsection{Inclination, frequency shifts, and self-absorption}

\begin{figure}[!htbp]
\centering
\includegraphics[width=0.90\linewidth]{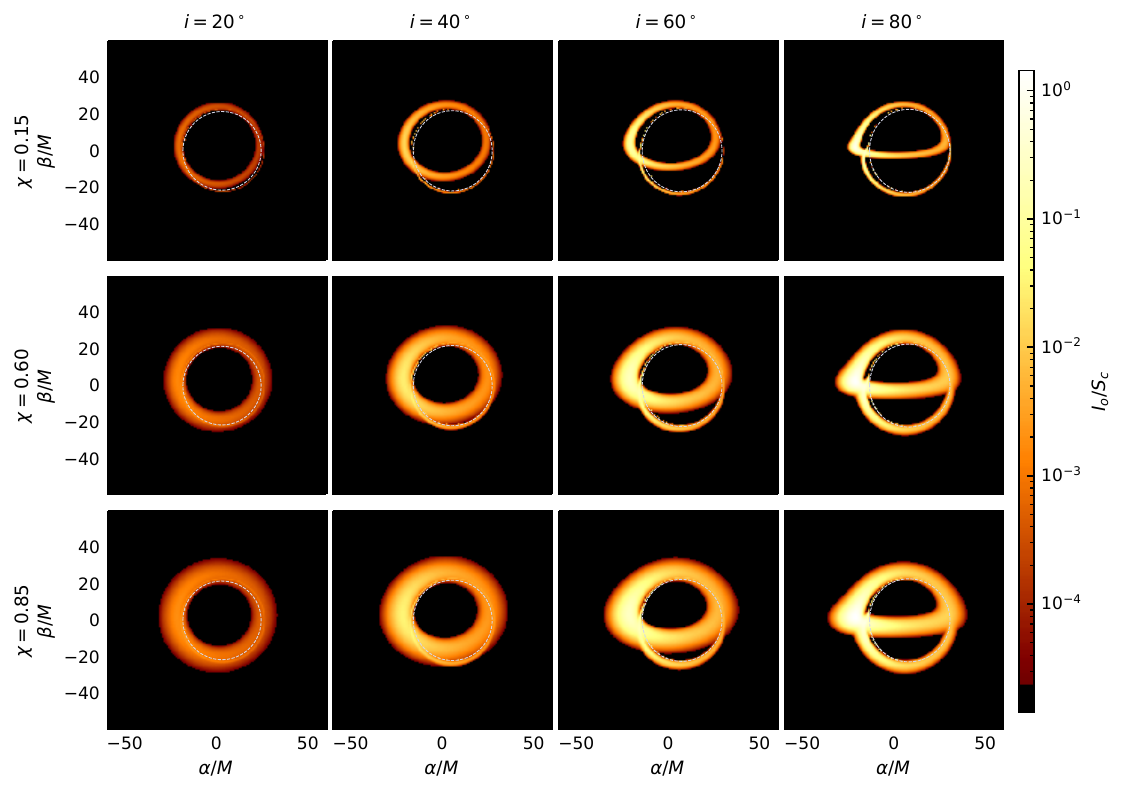}
\caption{Observed intensity at fixed $\widetilde BM=0.08$, $a/M=0.8$,
$r_c=4M$, $r_o=50M$, and $\tau_0=0.20$.  Rows give
$\chi=(0.15,0.60,0.85)$ and columns give
$i=(20^\circ,40^\circ,60^\circ,80^\circ)$.  The intensity normalization is
common to every panel.}
\label{fig:intensity-chii}
\end{figure}

At fixed geometry, Fig.~\ref{fig:intensity-chii} isolates viewing direction.
Low-inclination images are nearly annular.  Increasing $i$ produces a stronger
left--right contrast and separates near- and far-side contributions, while
larger $\chi$ increases the projected area and the range of emitting depths.

\begin{figure}[!htbp]
\centering
\includegraphics[width=0.90\linewidth]{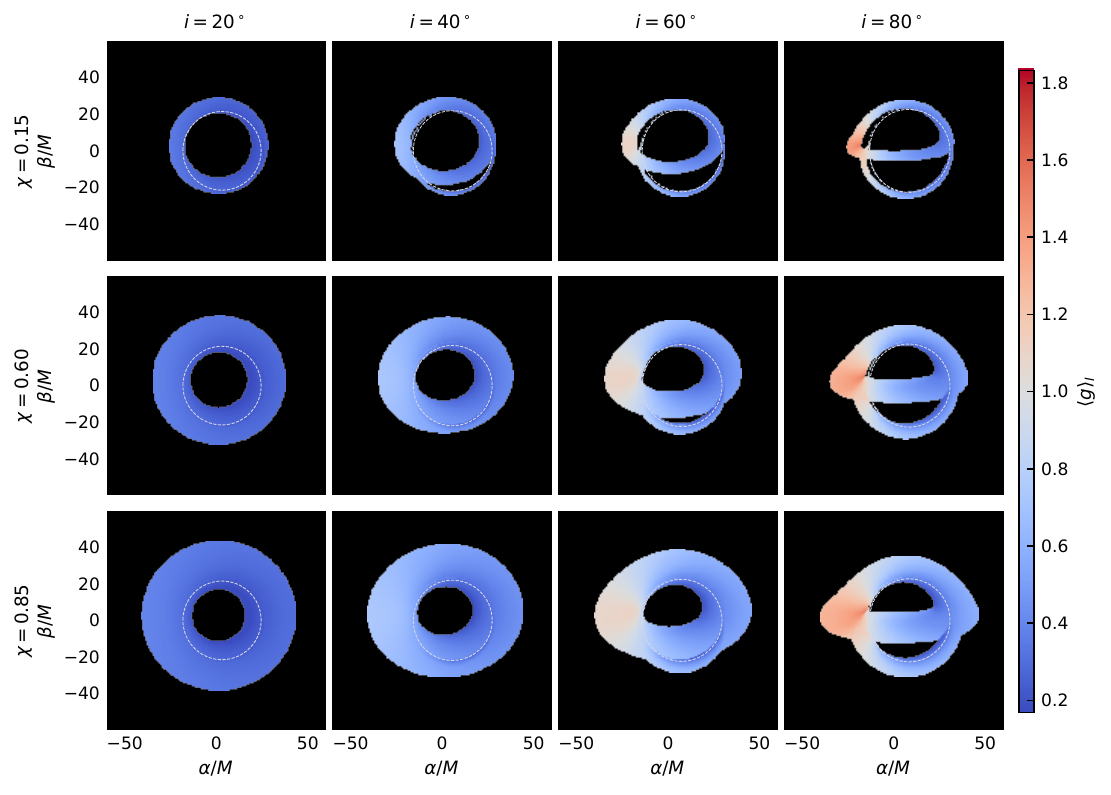}
\caption{Contribution-weighted frequency shift $\langle g\rangle_I$ for
the same filling--inclination grid as Fig.~\ref{fig:intensity-chii}.  Rows give
$\chi=(0.15,0.60,0.85)$ and columns give
$i=(20^\circ,40^\circ,60^\circ,80^\circ)$ at $r_o=50M$.  All panels share a diverging
scale centered on $g=1$; blue and red denote net redshift and blueshift, and
non-emitting directions are black.}
\label{fig:gstrip}
\end{figure}

The contribution-weighted frequency-shift maps in Fig.~\ref{fig:gstrip}
link the intensity asymmetry in Fig.~\ref{fig:intensity-chii} to the fluid
motion on the same filling--inclination grid.  The red--blue division rotates into a pronounced lateral
contrast as the line of sight approaches the equatorial plane, linking the
bright side in Fig.~\ref{fig:intensity-chii} to the fluid motion.  Increasing
$\chi$ broadens the region over which the same kinematic pattern is sampled;
it does not replace inclination as the main driver of the red--blue split.
The corresponding ray-class and optical-depth matrices are given in
Appendix~\ref{app:raydiagnostics}.

\begin{figure}[!htbp]
\centering
\includegraphics[width=0.90\linewidth]{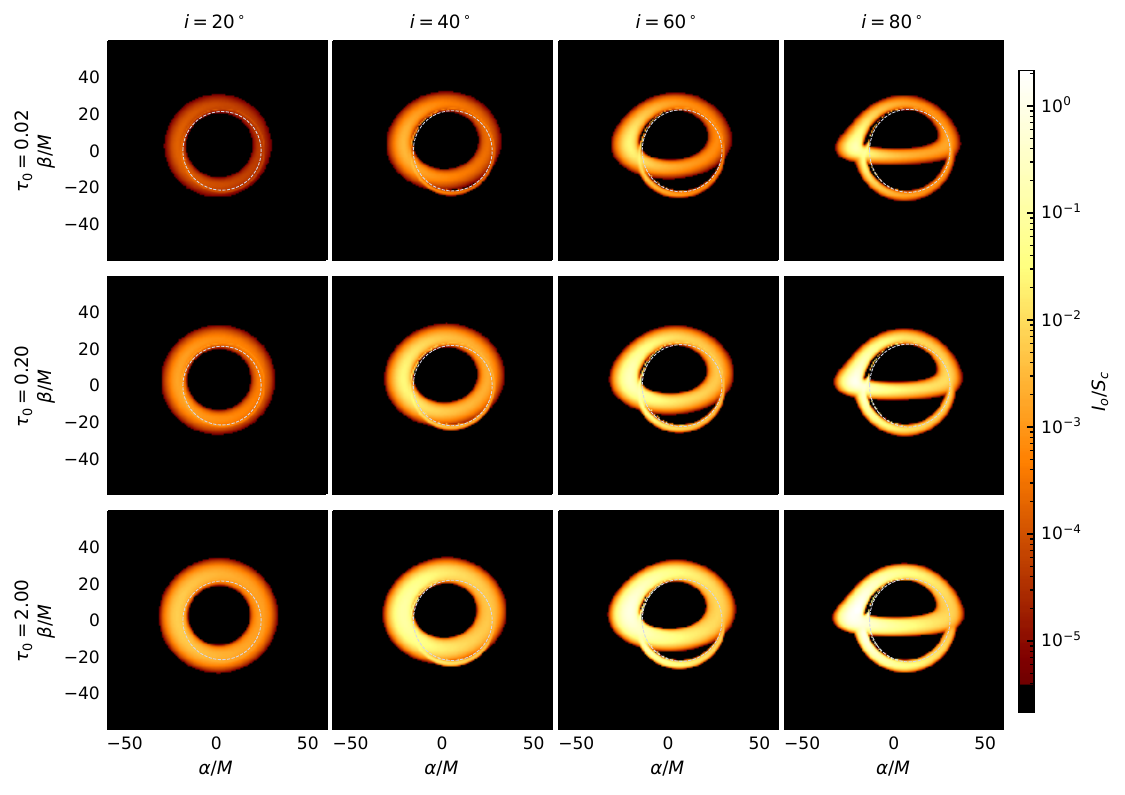}
\caption{Observed intensity for $\tau_0=(0.02,0.20,2.00)$ and
$i=(20^\circ,40^\circ,60^\circ,80^\circ)$, at $\chi=0.60$,
$a/M=0.8$, $r_c=4M$, $\widetilde BM=0.08$, and $r_o=50M$.  One absolute scale is used
for all panels.}
\label{fig:opacity-intensity}
\end{figure}

Figure~\ref{fig:opacity-intensity} then varies $\tau_0$ while holding the
equilibrium fixed.  At small $\tau_0$ the source is faint on the common
absolute scale; at larger $\tau_0$ the foreground approaches its local source
function and deeper contributions are suppressed, most visibly at high
inclination.  Because $j_e=\alpha_eS_e$, this changes emission and absorption
together and should be read as an optical-depth normalization survey, not as
pure absorption at fixed emissivity.

\FloatBarrier
\subsection{Equilibrium structure and rotational signatures}
\label{sec:spin-equilibrium}

Spin and rotation sense affect both the fluid motion and the equilibrium
free surface. The resulting image signatures depend on the spatial extent
and kinematics of each equilibrium. The pressure-center and common-center
rotation-sense sequences sample these coupled structural and frequency-shift
responses under different source constraints.
The field--filling sequence above uses $r_c=4M$, whereas the
pressure-center survey uses $\widetilde BM=0.04$ and $\chi=0.35$,
and the rotation-sense survey uses $\widetilde BM=0.05$ and $\chi=0.30$.
Each survey holds its own viewing and transfer parameters fixed.

\begin{figure}[!htbp]
\centering
\includegraphics[width=0.90\linewidth]{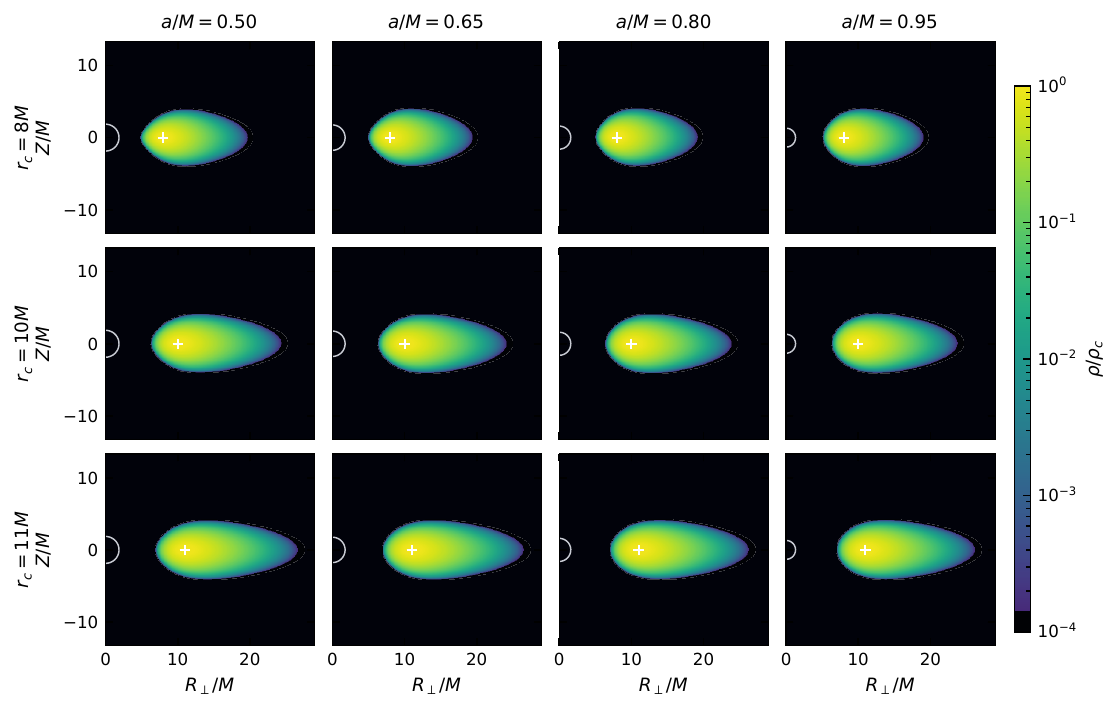}
\caption{Normalized equilibrium density at fixed $\chi=0.35$ and
$\widetilde BM=0.04$. Rows give $r_c/M=(8,10,11)$; columns give
$a/M=(0.50,0.65,0.80,0.95)$. A common coordinate window and density scale
are used.}
\label{fig:centerspin-density}
\end{figure}

Along the prograde branch, Fig.~\ref{fig:centerspin-density} varies the
pressure-center radius $r_c$ and spin $a$ at fixed $\chi$ and
$\widetilde B$. Moving the pressure
center outward substantially broadens the fluid region, whereas the change
across the selected spin columns is smoother. All twelve equilibria lie on
the same closure branch.

\begin{figure}[!htbp]
\centering
\includegraphics[width=0.90\linewidth]{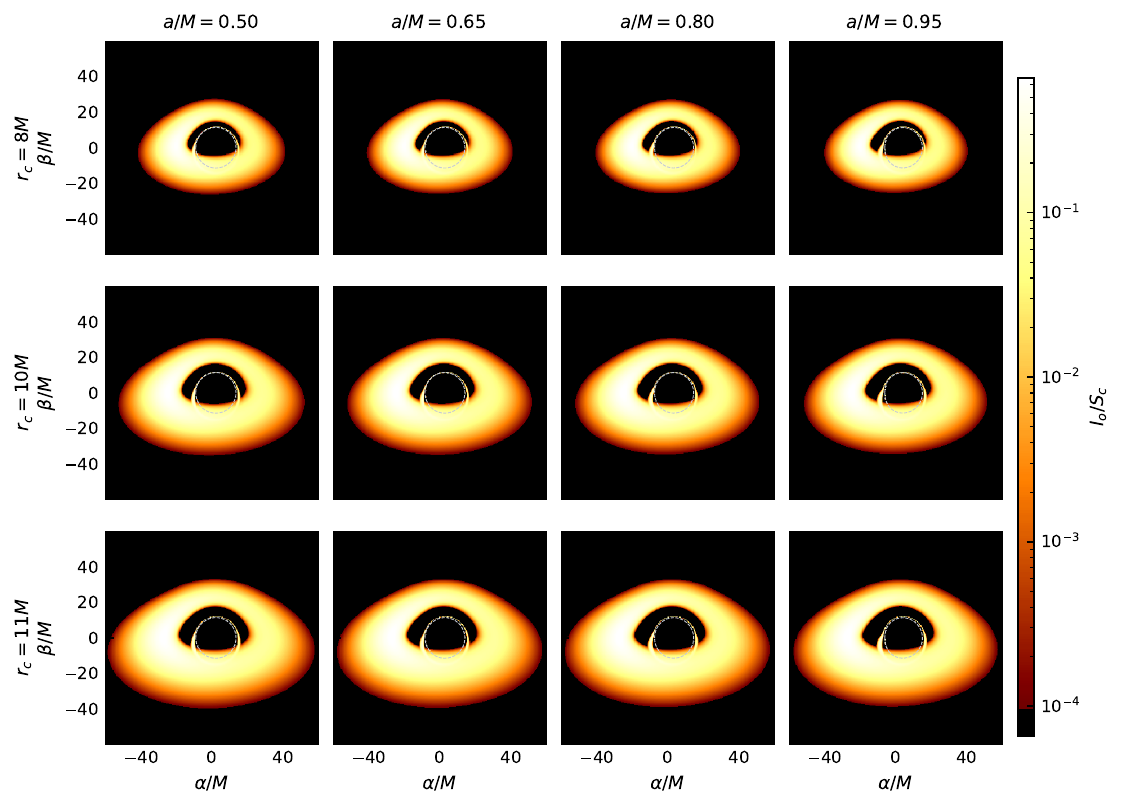}
\caption{Observed intensity for the equilibria in
Fig.~\ref{fig:centerspin-density}, at $r_o=50M$, $i=60^\circ$, and
$\tau_0=0.20$. Rows give $r_c/M=(8,10,11)$ and columns give
$a/M=(0.50,0.65,0.80,0.95)$. The $100^\circ$ field of view contains the
complete images in this sequence.}
\label{fig:centerspin-image}
\end{figure}

The intensity matrix in Fig.~\ref{fig:centerspin-image} shows the images of
these equilibria at $r_o=50M$.
Spin changes the near-side contrast and the position of the brightest arc,
but the outward displacement of the pressure center produces the larger
change in angular extent over the ranges shown.

\begin{figure}[!htbp]
\centering
\includegraphics[width=0.96\linewidth]{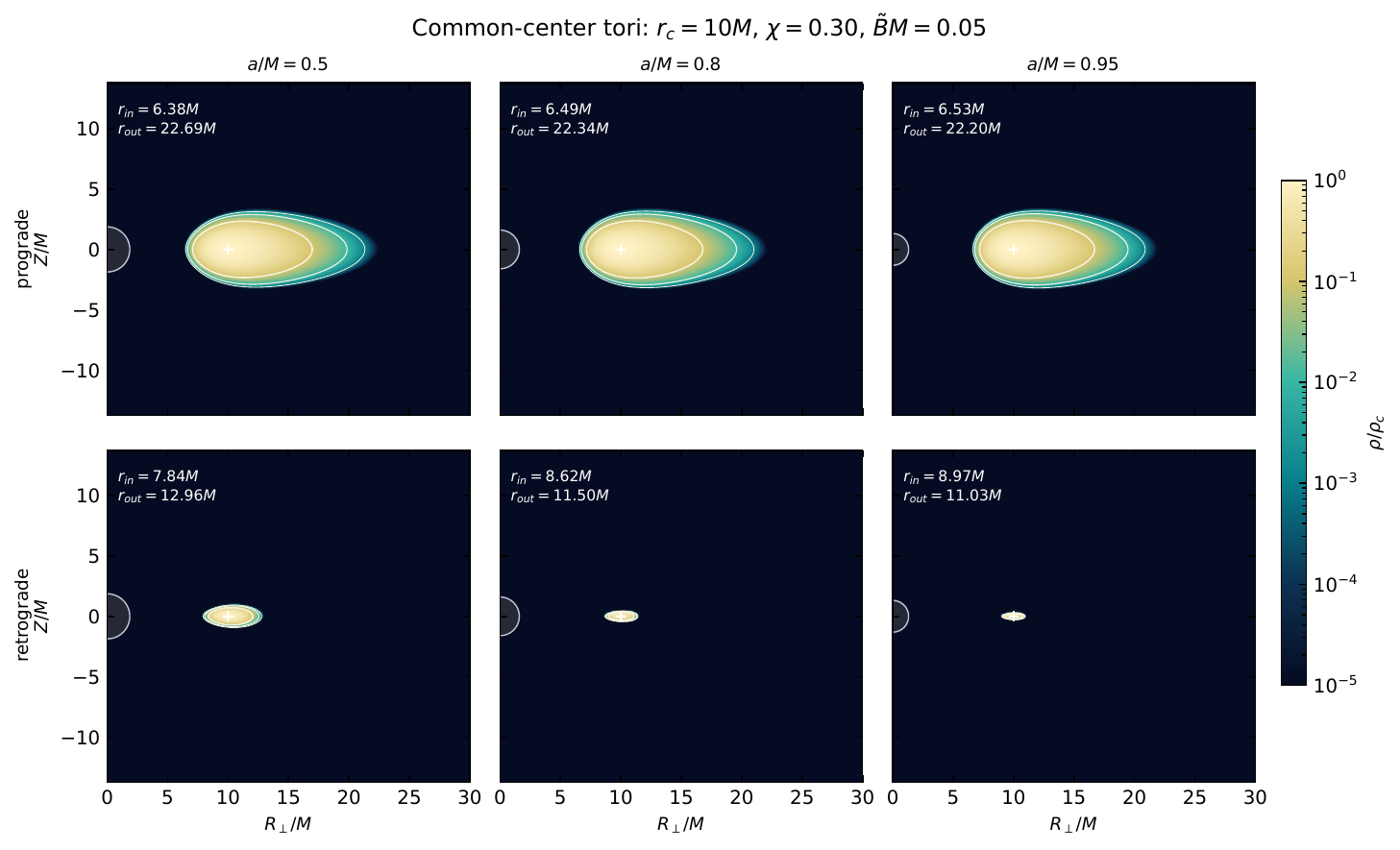}
\caption{Normalized density for the common-center sequence.  Columns give
$a/M=(0.5,0.8,0.95)$ and rows give prograde and retrograde rotation, at $r_c=10M$,
$\chi=0.30$, and $\widetilde BM=0.05$. The shared logarithmic scale exposes
the contraction of the retrograde lobe.}
\label{fig:spindensity}
\end{figure}

For the rotation-sense comparison we fix $r_c=10M$ and $\chi=0.30$,
with each signed $\ell$ set by its circular geodesic.  At $a/M=0.95$ and $\widetilde BM=0.05$, the selected prograde torus
spans $(r_{\rm in},r_{\rm out})=(6.533,22.204)M$, while the retrograde torus
spans only $(8.968,11.027)M$.  Equal center and filling do not imply equal
source volume. Figure~\ref{fig:spindensity} compares the density distributions
and free surfaces of these equilibria.

\begin{figure}[!htbp]
\centering
\includegraphics[width=0.98\linewidth]{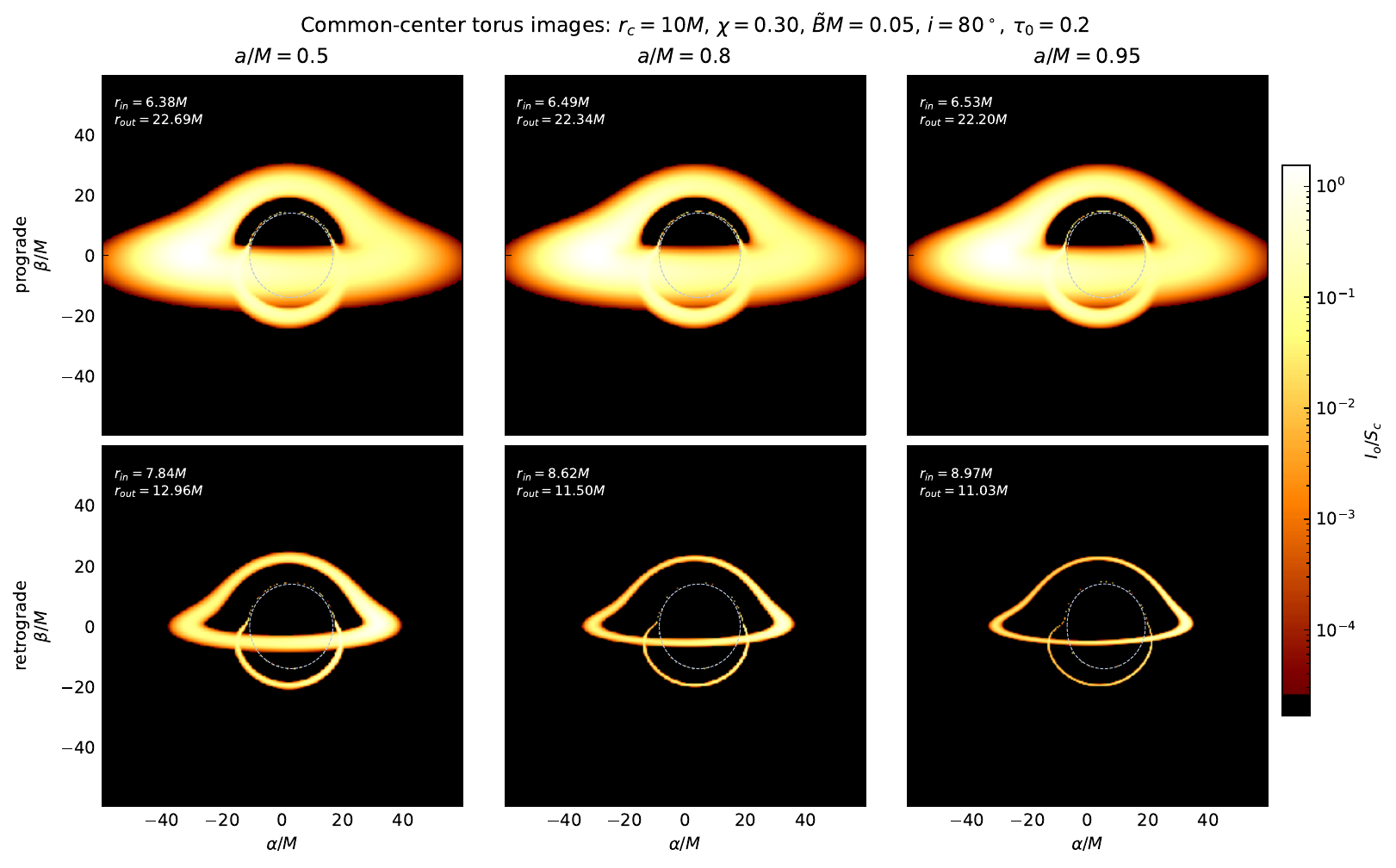}
\caption{Images of the common-center sequence at $r_c=10M$, $\chi=0.30$,
$\widetilde BM=0.05$, $r_o=50M$, $i=80^\circ$, and $\tau_0=0.20$.  Columns give spin
and rows give rotation sense.  A common absolute intensity scale is used;
dashed curves are the vacuum critical curves. The image grids are $384^2$
at $a/M=0.5,0.8$ and $768^2$ at $a/M=0.95$.}
\label{fig:spinimages}
\end{figure}

Figure~\ref{fig:spinimages} shows the corresponding high-inclination images.
The two rotation branches have opposite Doppler-bright sides, while the smaller angular
extent of the retrograde images also records their more compact equilibrium
sources. The common $r_c$ and $\chi$ specify the equilibrium family;
the two rotation senses have different free surfaces.

\begin{figure}[!htbp]
\centering
\includegraphics[width=0.90\linewidth]{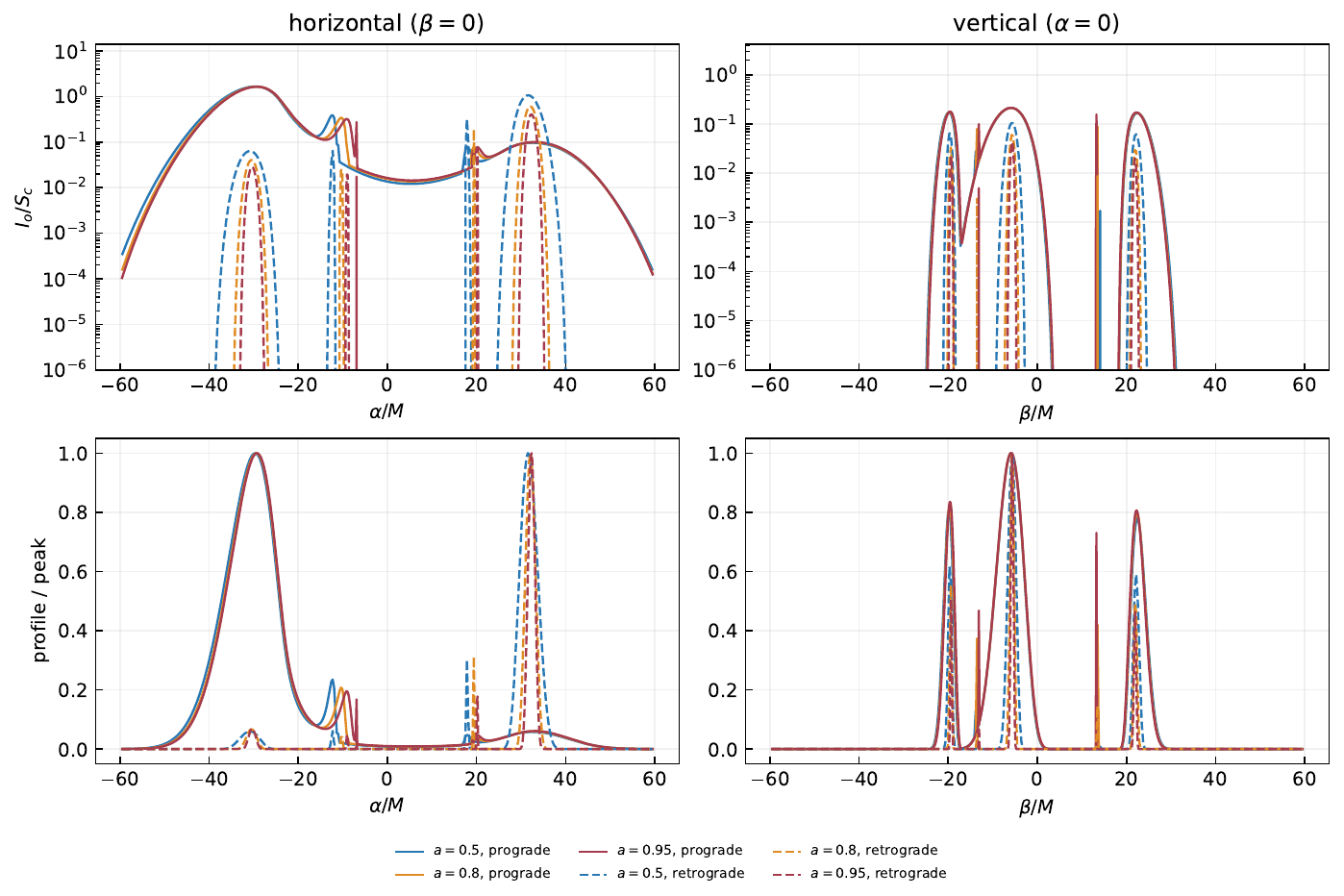}
\caption{Horizontal ($\beta=0$) and vertical ($\alpha=0$) cuts through
Fig.~\ref{fig:spinimages}.  The upper panels retain the common absolute
normalization; the lower panels normalize each cut to its own peak.  Solid
and dashed curves denote prograde and retrograde rotation. Each cut averages
the two central pixel rows or columns, with the grids stated in
Fig.~\ref{fig:spinimages}.}
\label{fig:spinprofiles}
\end{figure}

The one-dimensional cuts in Fig.~\ref{fig:spinprofiles} separate the broad
primary emission from narrow lensed features more clearly than the 2D maps.
On the common normalization, the screen-integrated intensity changes only
slightly along the prograde sequence but falls sharply along the retrograde
sequence as its lobe contracts.  The centroid changes sign when the flow is
reversed in the adopted screen convention.
Specifically, $\mathcal F$ changes from $0.13856$ to $0.13571$ along the
prograde sequence and from $0.02521$ to $0.00360$ along the retrograde
sequence as $a/M$ rises from $0.5$ to $0.95$.

An auxiliary equal-width comparison illustrates the sensitivity to the
chosen free surface. At $a/M=0.95$, reducing the prograde filling to
$\chi=0.01050$ matches its equatorial width to the retrograde value
$r_{\rm out}-r_{\rm in}=2.059M$ at $\chi=0.30$, while retaining
$r_c=10M$ and $\widetilde BM=0.05$. On a common $384^2$ image grid,
the prograde/retrograde integrated-intensity ratio then falls from $37.7$
for equal filling to $1.85$ for equal width. Matching this single coordinate
width does not match the full volume or thermodynamic profiles, so it does
not isolate a pure Doppler factor. The large equal-filling contrast is consequently sensitive to the
equilibrium boundary.

\begin{figure}[!htbp]
\centering
\includegraphics[width=0.90\linewidth]{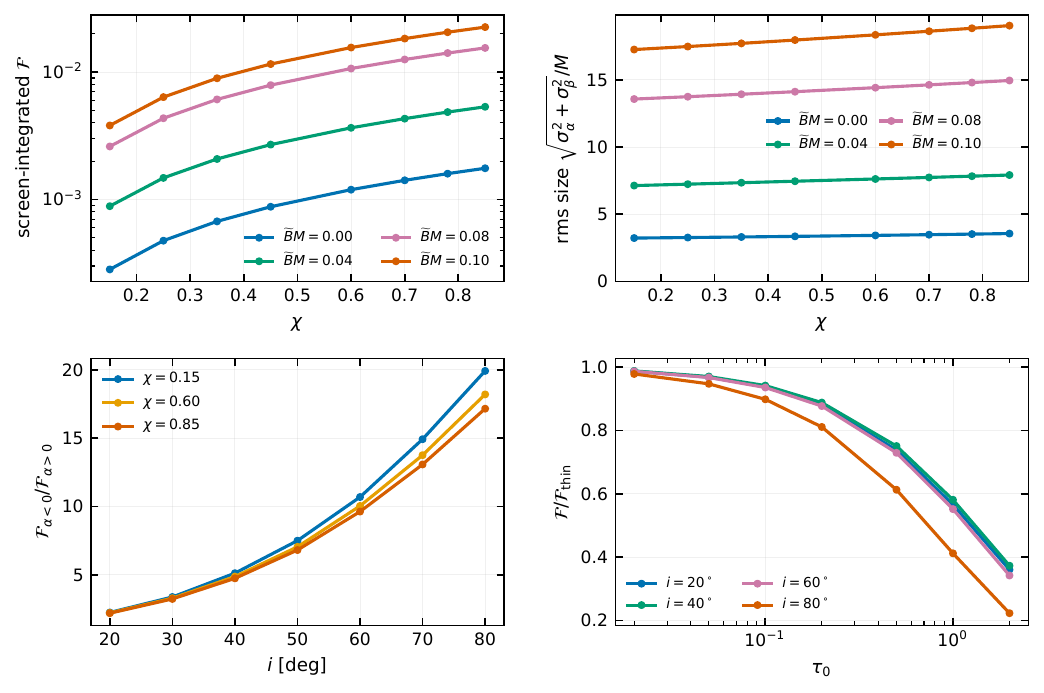}
\caption{Quantitative summary of the field--filling, inclination, and
optical-depth sequences in Figs.~\ref{fig:intensity-chib}--\ref{fig:opacity-intensity}.
Upper left: screen-integrated intensity versus filling.  Upper right: rms
image size.  Lower left: left/right flux ratio versus inclination.  Lower
right: attenuation relative to the optically thin integral versus $\tau_0$.
The scans use 8 filling values, 7 inclinations, and 7 approximately logarithmically spaced
optical-depth values; every marker is a separately ray-traced calculation on
a $224^2$ grid ($448^2$ for the Kerr field scan) at $r_o=50M$.
The line colors and legends identify the held or varied
parameter in each panel.}
\label{fig:observables}
\end{figure}

\medskip
\noindent
The integrated brightness, rms size, and lateral contrast distinguish the physical
responses across these image families. Figure~\ref{fig:observables} quantifies
the field--filling and viewing/transfer changes displayed in
Figs.~\ref{fig:intensity-chib}--\ref{fig:opacity-intensity}. Its upper panels
measure the integrated brightness and rms extent of the field--filling
images in Fig.~\ref{fig:intensity-chib}. The lower-left panel measures the
inclination-dependent brightness asymmetry of Fig.~\ref{fig:intensity-chii},
whose kinematic origin is illustrated by Fig.~\ref{fig:gstrip}. The
lower-right panel measures how much foreground absorption reduces the
intensity relative to the unabsorbed integral for the opacity sequence in
Fig.~\ref{fig:opacity-intensity}. These four panels summarize those sequences; the rotational signatures are
quantified separately by Figs.~\ref{fig:spinimages} and \ref{fig:spinprofiles}.
The denser parameter samples resolve their trends between the representative
image panels; each marker is an independent ray-tracing calculation.

The upper panels distinguish filling from the external-field response.
At fixed $\widetilde BM=0.08$, increasing $\chi$ from $0.15$ to $0.85$
raises $\mathcal F$ from $2.61\times10^{-3}$ to $1.55\times10^{-2}$,
a factor of $5.9$, while the rms size increases from $13.58M$ to $14.96M$,
only $10\%$. The growing torus therefore contributes substantially more
radiation without a comparable expansion of its bright image. The density
envelope in Fig.~\ref{fig:torusmatrix} and the intensity-weighted size measure
different parts of the source: low-density outer layers carry little weight
in the intensity moments. In contrast, at $\chi=0.60$, increasing
$\widetilde BM$ from $0$ to $0.10$ raises $\mathcal F$ from
$1.19\times10^{-3}$ to $1.56\times10^{-2}$ and the rms size from $3.43M$
to $18.36M$. Thus a large increase in brightness can accompany either a
modest size change along the filling sequence or a pronounced size change
along the field sequence. These are responses of the specified equilibrium
and finite-distance lens map, not a decomposition into independent source
and lensing contributions.

The lower-left panel quantifies the inclination dependence at fixed field.
For $\widetilde BM=0.08$ and $\chi=0.60$, changing $i$ from $20^\circ$ to
$80^\circ$ raises the left/right flux ratio from $2.23$ to $18.20$.
The frequency-shift maps in Fig.~\ref{fig:gstrip} locate the enhanced side,
while the ray and optical-depth maps in Appendix~\ref{app:raydiagnostics}
identify which paths cross foreground fluid. Inclination therefore changes
the contrast much more strongly than can be inferred from image size alone.

The lower-right panel separates self-absorption from total brightness.
At $i=80^\circ$, increasing $\tau_0$ from $0.02$ to $2.0$ lowers
$\mathcal F/\mathcal F_{\rm thin}$ from $0.978$ to $0.223$, while
$\mathcal F$ rises from $3.43\times10^{-3}$ to $7.84\times10^{-2}$.
An increasingly self-obscured source can consequently become brighter in
absolute terms in this model. The reason is that $\tau_0$ multiplies both
the absorption coefficient and the emissivity $j_e=\alpha_eS_e$; the
attenuation ratio measures the radiation lost relative to the corresponding
unabsorbed source, not relative to a fixed-emissivity image.

A limited source-function test replaces $q^4$ by $q^{n_{\rm em}}$, with
$n_{\rm em}=2,4,6$,
in the fiducial $i=80^\circ$ model on a $192^2$ grid. At $\tau_0=2$,
the attenuation ratio ranges from $0.20$ to $0.25$, and the absolute
intensity still rises with $\tau_0$ for each exponent. The absolute
intensities at that optical depth differ by a factor of $2.2$ across the
three prescriptions. The coexistence of increasing brightness and stronger
self-absorption thus persists in this test, while its numerical amplitude
depends on the assumed concentration of the source function.

The different brightness, size, and asymmetry responses across these
families reflect the combined effects of equilibrium structure, fluid
motion, and radiative transfer.

\FloatBarrier
\section{Discussion and conclusions}

We have calculated gray images of stationary equilibrium tori in the
neutral KBR spacetime, integrating emission and absorption through
their finite volumes to an observer at $r_o=50M$. The field and filling
sequences produce different relations between brightness and image size.
Increasing $\chi$ from $0.15$ to $0.85$ at $\widetilde BM=0.08$ raises
integrated intensity by a factor of $5.9$ but increases rms size by only
$10\%$. At $\chi=0.60$, the rms size instead grows by a factor of $5.35$
between $\widetilde BM=0$ and $0.10$. The equilibrium envelope and the
bright-image extent therefore need not vary in proportion. The field
response combines changes in the fluid potential and the lens map,
whereas filling changes the free surface within a fixed spacetime.

Inclination enhances lateral brightness contrast, while self-absorption
changes the emitting depths visible to the observer. At $i=80^\circ$,
increasing the optical-depth normalization from $0.02$ to $2$ lowers
the received fraction of the unabsorbed intensity from $0.98$ to $0.22$,
even as absolute brightness rises. This coexistence follows from
$j_e=\alpha_eS_e$: emission and absorption scale together in the gray
prescription. The attenuation ratio and absolute intensity consequently
measure different aspects of the source. Optically thin KBR disk models
\cite{ThinKNBR,Synchrotron} omit this attenuation through the emitting volume.

Rotation sense affects the source structure as well as the bright side.
At common pressure center and filling, increasing $a/M$ from $0.5$ to
$0.95$ dims the retrograde sequence by a factor of about $7$, whereas
prograde brightness changes by only $2\%$. The retrograde dimming
accompanies contraction of its equilibrium lobe. Motion-dependent photon
escape and frequency shifts are quantified for Kerr geodesic emitters
in Ref.~\cite{SuKerrEmission}; the torus comparison also changes the spatial
distribution of emission. Its rotational brightness contrast thus contains
both kinematic and equilibrium-structure effects, rather than a pure
Doppler factor.

These results apply to stationary, neutral test fluids with the specified
gray source function and finite observer. Internal magnetic stresses
\cite{Komissarov,HouMagnetofluids}, plasma emission prescriptions
\cite{StraubIonTorus,VincentTorus,Dexter}, and polarized transfer
\cite{IPole,Coport,HouPolar} require independent physical assumptions for
observational predictions. Appendix~\ref{app:observer-radius} quantifies the
remaining observer-radius dependence. Within the investigated families,
brightness, size, and asymmetry exhibit different responses to the source
and transfer parameters. Their joint interpretation must account for
the equilibrium emitting region as well as the propagation of null rays.

\section*{Acknowledgements}
\begin{sloppypar}
This work was supported by the National Natural Science Foundation of China under Grant No.~12305070, and the Basic Research Program of Shanxi Province under Grant Nos.~202303021222018 and 202303021221033.
\end{sloppypar}

\FloatBarrier
\appendix
\numberwithin{figure}{section}
\section{Auxiliary vacuum and circular-source lens maps}
\label{app:vacuum}

\begin{figure}[!htbp]
\centering\includegraphics[width=0.86\linewidth]{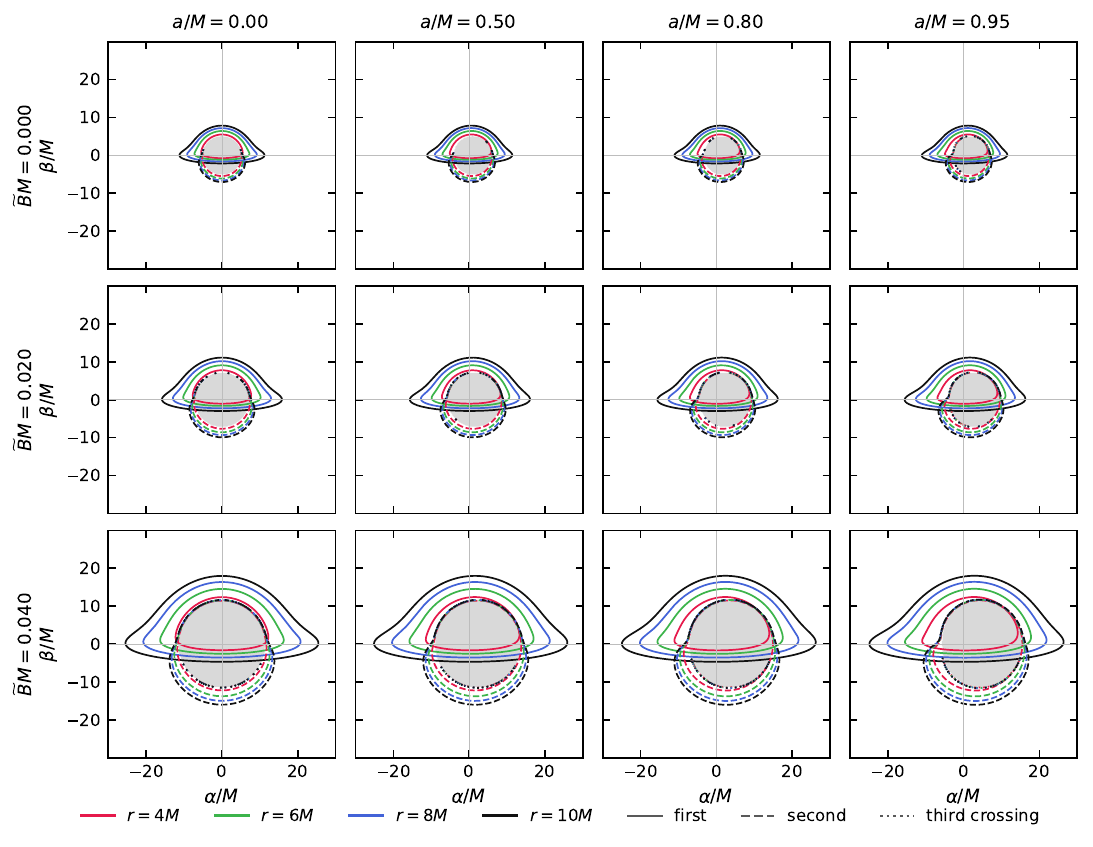}
\caption{Equatorial circular-source images at $r_o=50M$, $i=80^\circ$.
Rows give $\widetilde BM=(0,0.02,0.04)$ and columns give
$a/M=(0,0.50,0.80,0.95)$. Colors identify $r/M=(4,6,8,10)$; solid, dashed,
and dotted curves denote successive equatorial crossings. Gray denotes
captured directions.}
\label{fig:ringmatrix}
\par\vspace{6pt}
\includegraphics[width=0.86\linewidth]{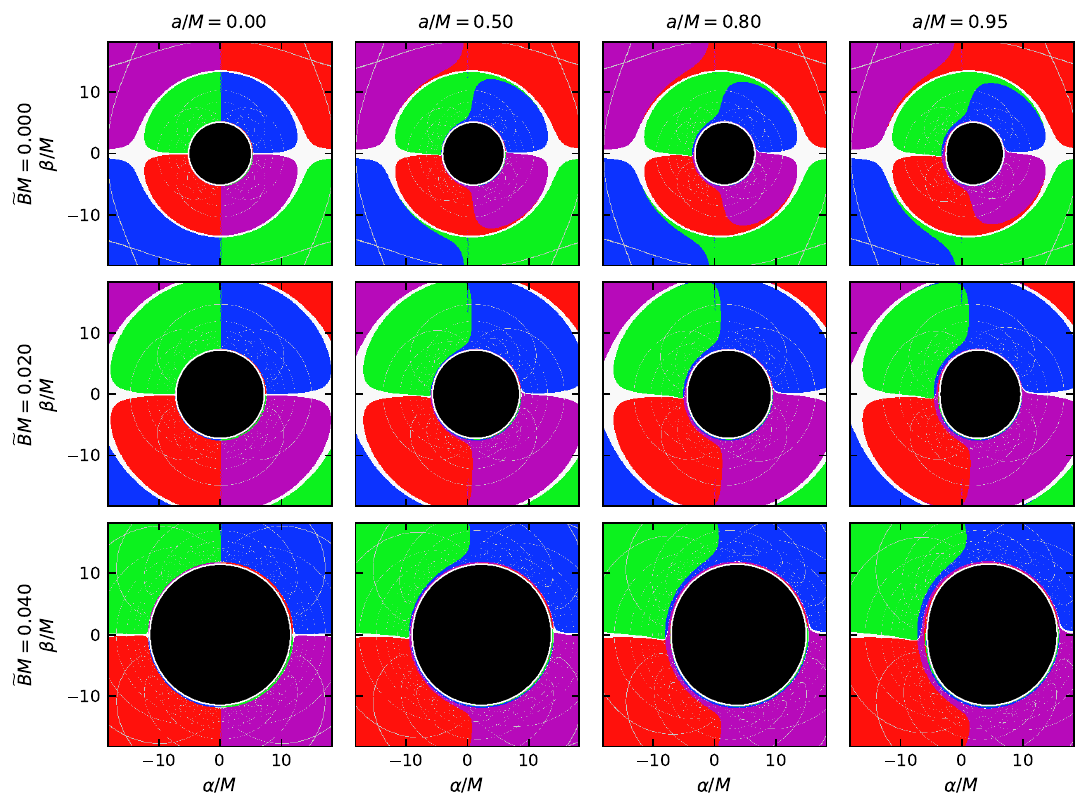}
\caption{Lensed celestial sphere at $r_s=100M$, using the spacetime and
observer grid of Fig.~\ref{fig:ringmatrix} and a $40^\circ$ field of view.
Colors identify source-texture quadrants in the tangent frame about the
antipodal optical direction; black denotes capture and the light curves
are mapped source-coordinate grid lines.}
\label{fig:skymatrix}
\end{figure}

Prescribed circular emitters and a celestial sphere isolate null propagation
without torus absorption. Figure~\ref{fig:ringmatrix} shows the first three
equatorial crossings; its source radii and field samples differ from the
torus grid in Figs.~\ref{fig:torusmatrix} and \ref{fig:intensity-chib}.

\FloatBarrier

At the same observer radius, the finite celestial-source map displays the
field- and spin-dependent distortion of a sphere at $r_s=100M$. Grid folding
close to the capture boundary marks strongly bent paths. The representative
trajectories then illustrate that bending at $\widetilde BM=0$ and $0.10$;
their meridional projections use common axes.

\begin{figure}[H]
\centering
\includegraphics[width=0.94\linewidth]{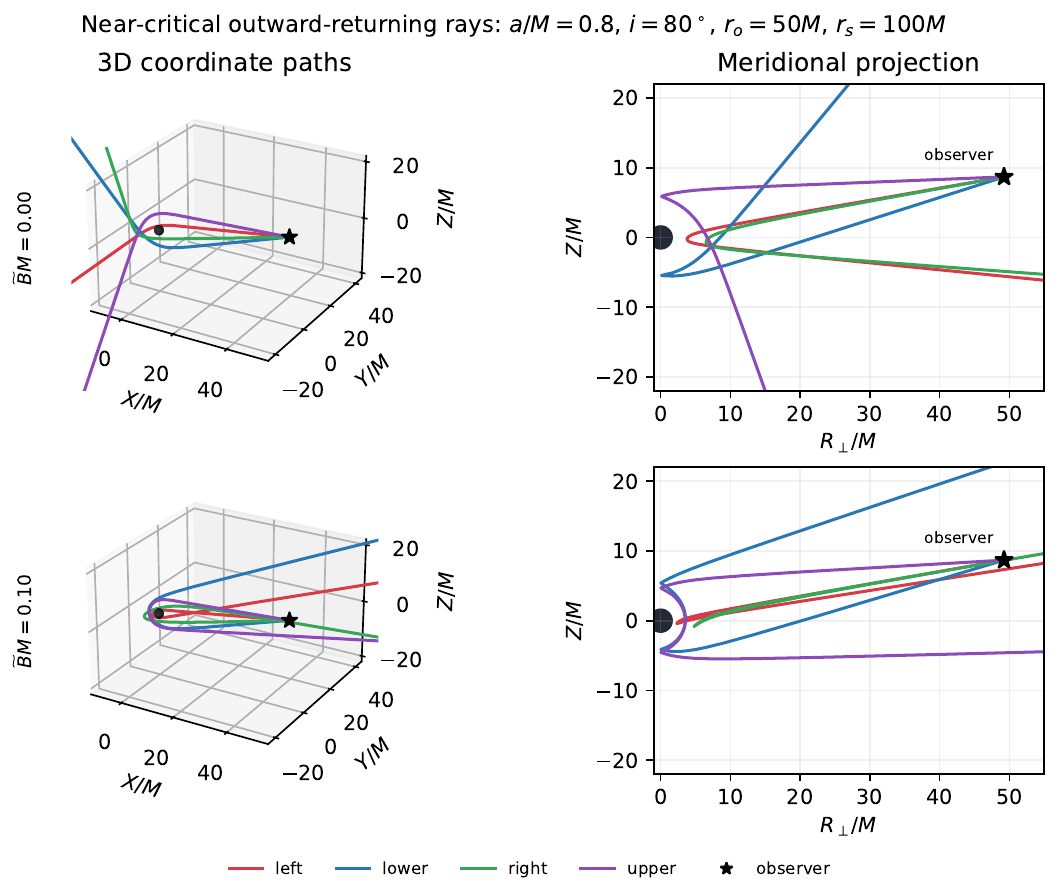}
\caption{Four representative outward-returning null rays selected just outside the capture boundary for $a/M=0.8$, $i=80^\circ$, $r_o=50M$, and $r_s=100M$.  Rows give $\widetilde BM=0$ and $0.10$; the left column shows three-dimensional coordinate paths and the right column their meridional projections.  Common, cropped coordinate windows emphasize the strongly bent near region; the integrations themselves continue to the source sphere at $r_s=100M$.  Colors identify the four source-sphere directions, and the black star marks the observer.}
\label{fig:trajectories}
\end{figure}

\section{Ray classification and optical-depth maps}
\label{app:raydiagnostics}
\label{app:optdepth}

\begin{figure}[!htbp]
\centering\includegraphics[width=0.90\linewidth]{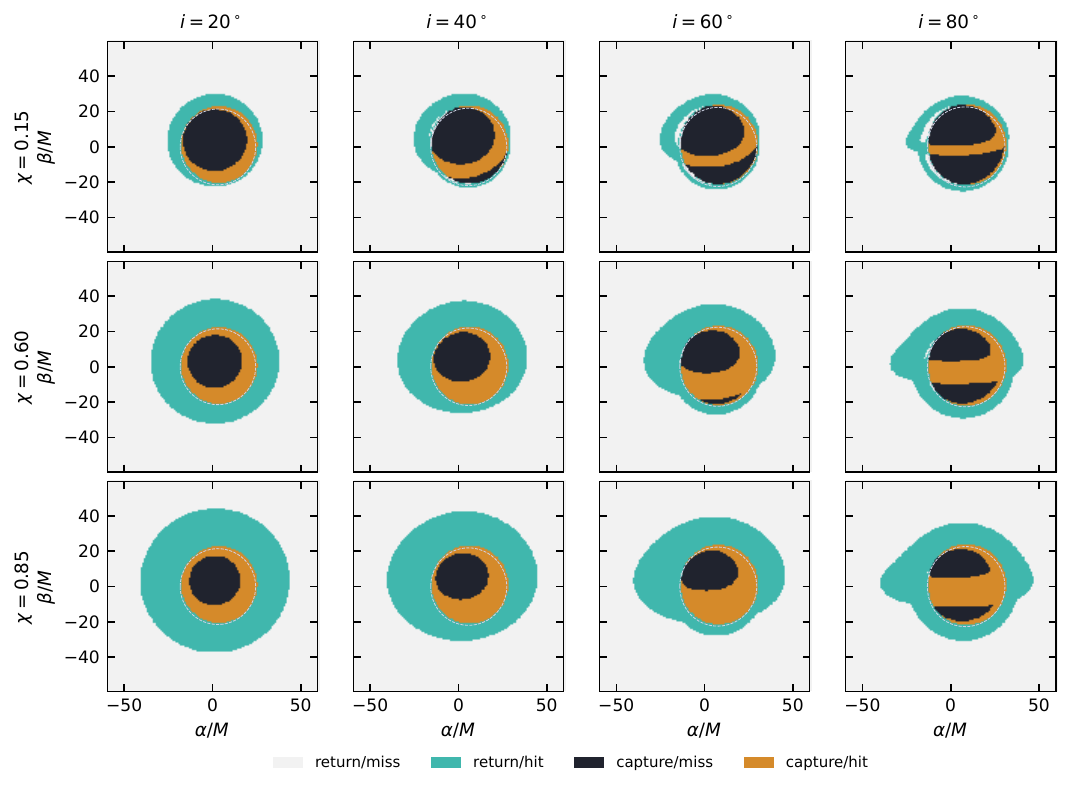}
\caption{Ray classification for the $r_o=50M$ models shown in Fig.~\ref{fig:intensity-chii}. Rows give $\chi=(0.15,0.60,0.85)$ and columns give $i=(20^\circ,40^\circ,60^\circ,80^\circ)$. The shared legend distinguishes horizon capture from return to $r=r_o$ and records whether a ray intersects the torus.}
\label{fig:classmatrix}
\end{figure}

Figure~\ref{fig:classmatrix} separates horizon capture from return to the finite outer boundary and records whether each ray crosses the torus. Repetition along the filling axis is physical: capture or return is source-independent, while the torus-intersection label changes with the source boundary.

\begin{figure}[!htbp]
\centering\includegraphics[width=0.90\linewidth]{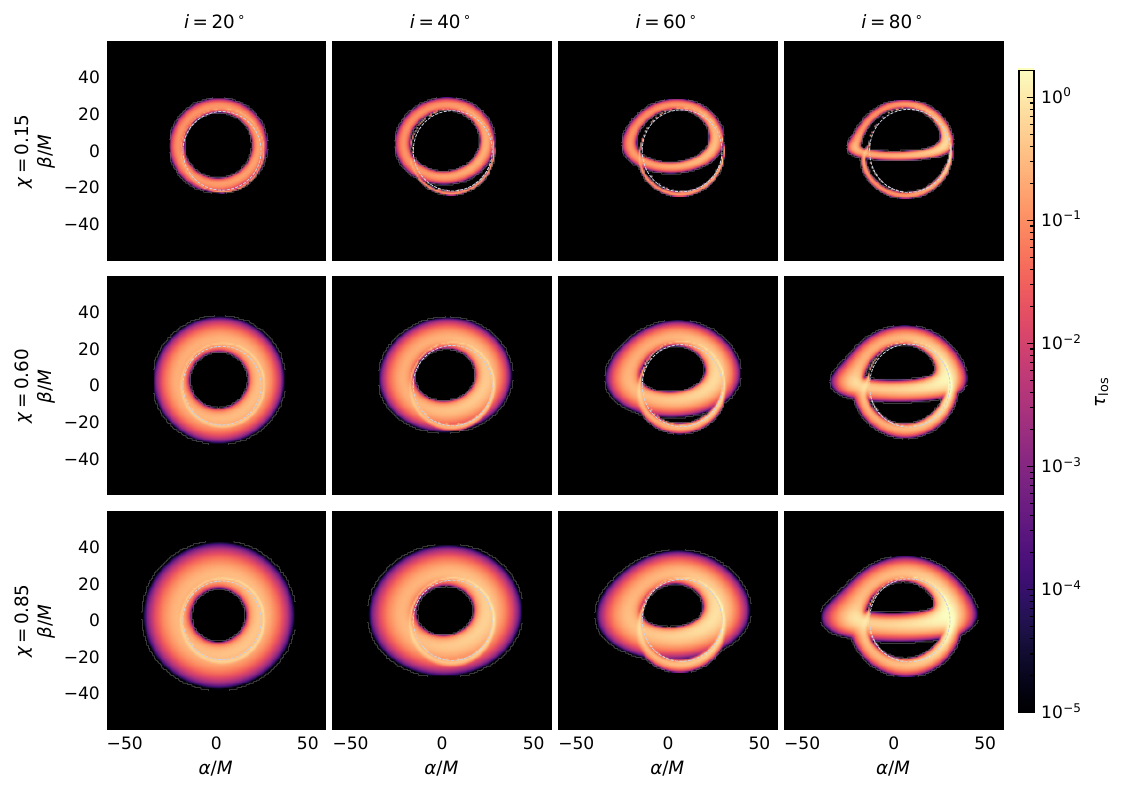}
\caption{Integrated optical depth for the same rays as Fig.~\ref{fig:intensity-chii}. Rows give $\chi=(0.15,0.60,0.85)$, columns give increasing inclination, and all panels use one logarithmic scale.}
\label{fig:taumatrix}
\end{figure}

Figure~\ref{fig:taumatrix} gives the integrated optical-depth maps for the same filling--inclination sequence. They locate the columns most susceptible to foreground attenuation. Larger filling and higher inclination enlarge the projected high-depth regions, although the emergent intensity also depends on the source function and frequency shift.

\FloatBarrier
\section{Observer-radius sensitivity}
\label{app:observer-radius}

\begin{figure}[!htbp]
\centering
\includegraphics[width=0.98\linewidth]{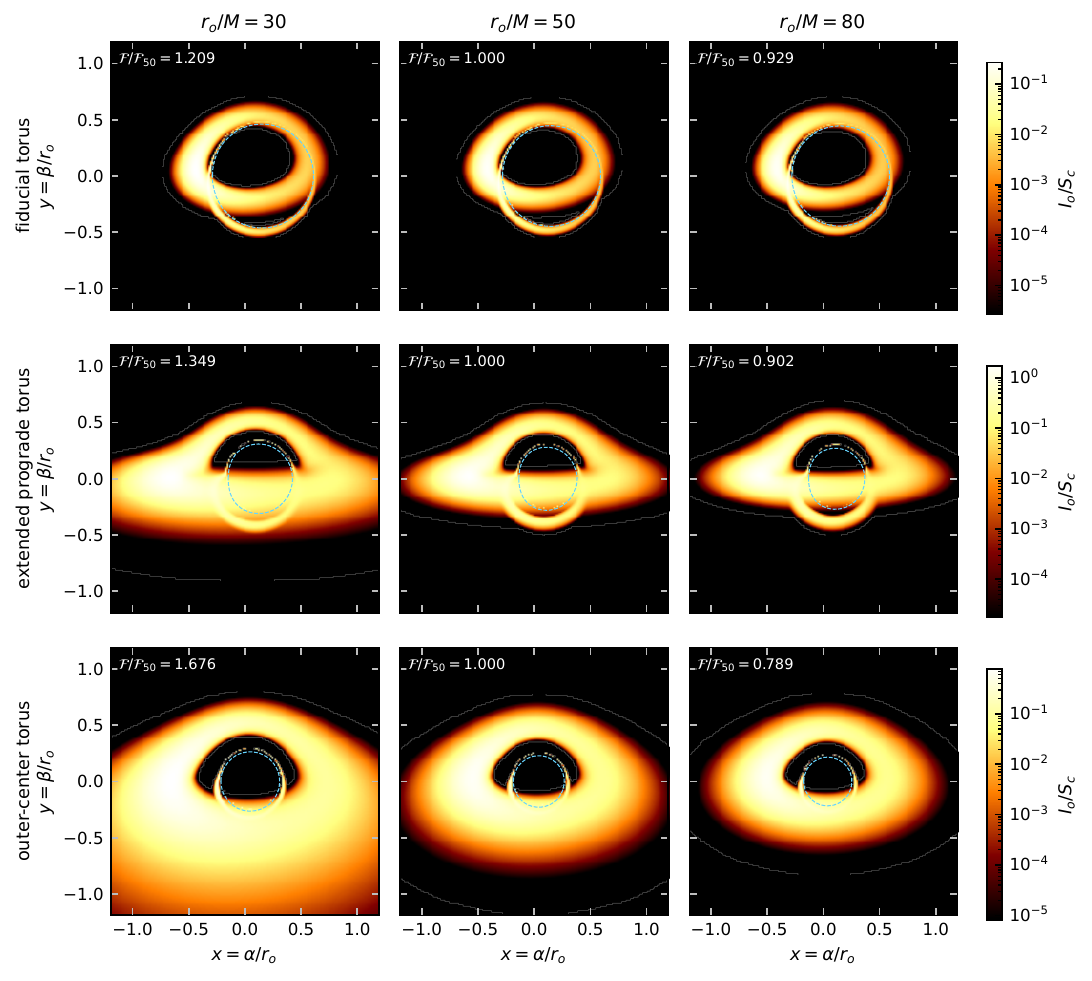}
\caption{Finite-observer-radius comparison at $r_o/M=(30,50,80)$. The upper
row is the fiducial model $(a/M,\widetilde BM,r_c/M,\chi,i)=(0.8,0.08,4,0.60,
60^\circ)$; the middle row is the extended prograde model
$(0.95,0.05,10,0.30,80^\circ)$ with $r_{\rm out}=22.204M$; the lower row is
the outer-center model $(0.50,0.04,11,0.35,60^\circ)$ with
$r_{\rm out}=27.592M$. Intensities are
shown on one logarithmic scale within each row, dashed curves are the vacuum
critical curves, and the inset numbers give the screen-integrated intensity
relative to the $r_o=50M$ calculation.}
\label{fig:observer-radius-images}
\end{figure}

We quantify the observer-radius dependence of the baseline $r_o=50M$
images by repeating the ray tracing at
$r_o/M=(30,40,50,60,70,80)$ while holding the local $100^\circ$ gnomonic field of view,
source model, inclination, opacity normalization, and numerical step control
fixed. Figure~\ref{fig:observer-radius-images} compares the fiducial torus, an
extended high-spin prograde torus, and the largest equilibrium in the
pressure-center sequence. The
coordinates $x=\alpha/r_o$ and $y=\beta/r_o$ are local angular screen
coordinates, so the changing outline is not hidden by rescaling with $r_o$.

We measure the angular-size proxy
$\psi_{\rm rms}=\arctan\sqrt{\sigma_x^2+\sigma_y^2}$, using the
intensity-weighted gnomonic variances
$\sigma_x=\sigma_\alpha/r_o$ and $\sigma_y=\sigma_\beta/r_o$, and the median critical-curve angular
radius $\arctan\sqrt{x^2+y^2}$; these avoid the explicit $r_o$ scaling of
$\alpha$ and $\beta$.
For the fiducial torus, moving the observer from $50M$ to $80M$ changes the
angular rms size by only $0.20\%$, the median critical-curve radius by
$0.38\%$, and the left/right flux ratio by $0.33\%$. For the extended torus,
the same displacement changes those quantities by
$3.00\%$, $3.07\%$, and $0.52\%$, respectively. In contrast, the
screen-integrated intensity changes by $7.1\%$ and $9.8\%$ in the two models.
For the largest pressure-center model, with $r_{\rm out}=27.592M$, the
$50M$--$80M$ changes are $6.23\%$ in angular size, $5.20\%$ in critical-curve
radius, $0.58\%$ in left/right ratio, and $21.09\%$ in integrated intensity.
The observer is only at $1.09r_{\rm out}$ when $r_o=30M$, compared with
$1.81r_{\rm out}$ at $50M$. The extended-source survey therefore retains
appreciable observer-radius dependence in both size and intensity.
Across all three cases, the asymmetry is substantially more stable than
the absolute intensity. The $r_o=50M$ comparison is therefore a specified
finite-distance experiment, with the residual dependence reported here.

\begin{figure}[!htbp]
\centering
\includegraphics[width=0.94\linewidth]{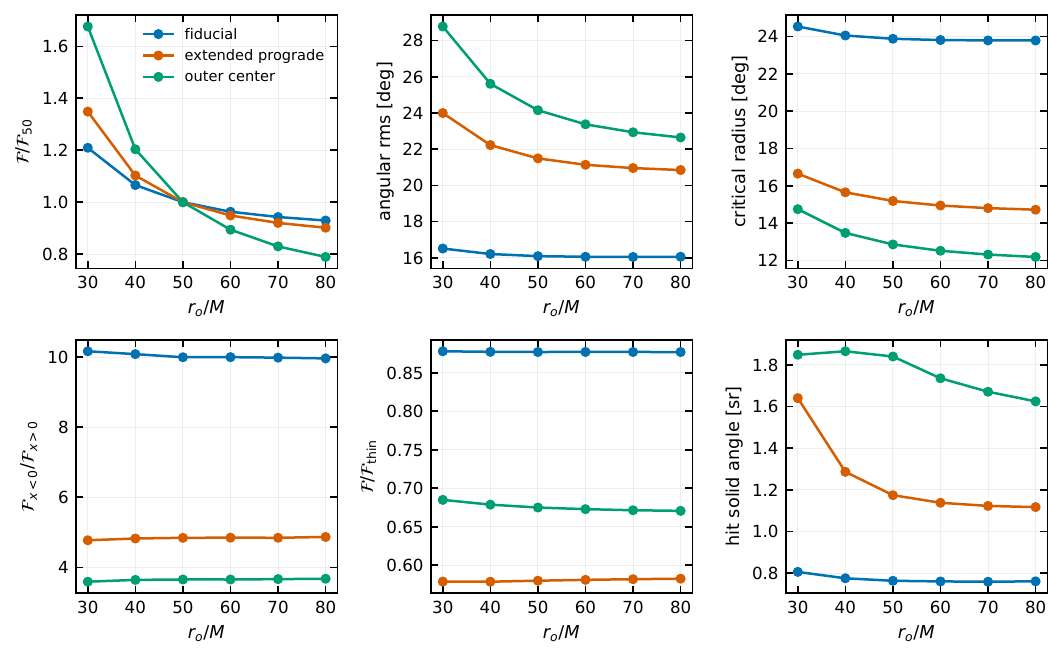}
\caption{Dependence of image properties on the observer radius for the three rows of
Fig.~\ref{fig:observer-radius-images}. From left to right and top to bottom:
relative integrated intensity, angular rms size, median critical-curve
radius, left/right flux ratio, attenuation relative to the optically thin
integral, and torus-hit solid angle. Each sequence samples
$r_o/M=(30,40,50,60,70,80)$, whereas the image comparison retains the three
representative radii $30M$, $50M$, and $80M$. Lines connect independently ray-traced
calculations only as visual guides.}
\label{fig:observer-radius-metrics}
\end{figure}

Figure~\ref{fig:observer-radius-metrics} gives the corresponding scalar
radius dependence. The integrated-intensity and angular-size curves
flatten gradually over the sampled interval. The larger changes between $30M$ and $50M$ are
especially important for the extended torus: because $r_o/r_{\rm out}=1.35$
at $30M$ but $2.25$ at $50M$, the former position samples a visibly more local
view. All integrals here refer to the fixed $100^\circ$ aperture; the extended
sources reach its edges most visibly at $30M$. A $192^2$ grid is used for
these supplementary calculations. Comparing the nine image models at
$r_o/M=(30,50,80)$ against a
$128^2$ grid changes the integrated intensities by at most $0.28\%$ and the
angular rms sizes by at most $0.17\%$. All rays reach a terminal state, and
the largest first-integral residual over the six observer radii remains
below $2.8\times10^{-5}$.

\FloatBarrier

\begingroup
\small
\interlinepenalty=10000
\bibliographystyle{unsrtnat}
\bibliography{references}

@article{OldSolution,
  author = {J. Podolsky and H. Ovcharenko},
  title = {{Kerr black hole in a uniform Bertotti--Robinson magnetic field: An exact solution}},
  year = {2025},
  journal = {Phys. Rev. Lett.},
  volume = {135},
  pages = {181401},
  eprint = {2507.05199},
  archivePrefix = {arXiv},
  primaryClass = {gr-qc},
  note = {\href{https://arxiv.org/abs/2507.05199}{arXiv:2507.05199}},
  doi = {10.1103/rfgv-ybz5}
}

@misc{Genuine,
  author = {H. Ovcharenko and J. Podolsky},
  title = {{Kerr--Newman and genuine Kerr black holes in the Bertotti--Robinson magnetic field}},
  year = {2026},
  eprint = {2608.21672},
  archivePrefix = {arXiv},
  primaryClass = {gr-qc},
  note = {\href{https://arxiv.org/abs/2608.21672}{arXiv:2608.21672}}
}

@misc{Thermo,
  author = {D. Kubiznak and H. Ovcharenko and J. Podolsky},
  title = {{Thermodynamics of Genuine Kerr--Bertotti--Robinson Black Holes}},
  year = {2026},
  eprint = {2608.22292},
  archivePrefix = {arXiv},
  primaryClass = {gr-qc},
  note = {\href{https://arxiv.org/abs/2608.22292}{arXiv:2608.22292}}
}

@article{ShadowOld,
  author = {X. Wang and Y. Hou and X. Wan and M. Guo and B. Chen},
  title = {{Geodesics and shadows in the Kerr--Bertotti--Robinson black hole spacetime}},
  year = {2026},
  journal = {JCAP},
  volume = {02},
  pages = {050},
  eprint = {2507.22494},
  archivePrefix = {arXiv},
  primaryClass = {gr-qc},
  note = {\href{https://arxiv.org/abs/2507.22494}{arXiv:2507.22494}}
}

@article{OpticalOld,
  author = {X.-X. Zeng and C.-Y. Yang and H. Yu},
  title = {{Optical characteristics of the Kerr--Bertotti--Robinson black hole}},
  year = {2025},
  journal = {Eur. Phys. J. C},
  volume = {85},
  pages = {1242},
  eprint = {2508.03020},
  archivePrefix = {arXiv},
  primaryClass = {gr-qc},
  note = {\href{https://arxiv.org/abs/2508.03020}{arXiv:2508.03020}}
}

@article{PhotonRing,
  author = {X. Wan and Z. Zhang and F.-S. Wei and Y. Hou and B. Chen},
  title = {{Critical behavior of photon rings in Kerr--Bertotti--Robinson spacetime}},
  year = {2026},
  journal = {Phys. Rev. D},
  volume = {114},
  pages = {044005},
  eprint = {2603.25049},
  archivePrefix = {arXiv},
  primaryClass = {gr-qc},
  note = {\href{https://arxiv.org/abs/2603.25049}{arXiv:2603.25049}}
}

@misc{ThinKNBR,
  author = {X. Wan and H. Wang and Z. Zhang and Z. Zhang and B. Chen},
  title = {{Shadows and thin-disk images of Kerr--Newman black holes in a Bertotti--Robinson magnetic field}},
  year = {2026},
  eprint = {2609.06648},
  archivePrefix = {arXiv},
  primaryClass = {gr-qc},
  note = {\href{https://arxiv.org/abs/2609.06648}{arXiv:2609.06648}}
}

@article{Synchrotron,
  author = {Z.-Y. Zhang and X.-Q. Li and H.-P. Yan and X.-J. Yue},
  title = {{Optical appearance of the Kerr--Bertotti--Robinson black hole with a magnetically driven synchrotron emissivity model}},
  year = {2026},
  journal = {Eur. Phys. J. C},
  volume = {86},
  pages = {945},
  eprint = {2605.09593},
  doi = {10.1140/epjc/s10052-026-16174-1},
  archivePrefix = {arXiv},
  primaryClass = {astro-ph.HE},
  note = {\href{https://arxiv.org/abs/2605.09593}{arXiv:2605.09593}}
}

@article{Fishbone,
  author = {L. G. Fishbone and V. Moncrief},
  title = {{Relativistic fluid disks in orbit around Kerr black holes}},
  year = {1976},
  journal = {Astrophys. J.},
  volume = {207},
  pages = {962},
  url = {https://articles.adsabs.harvard.edu/pdf/1976ApJ...207..962F}
}

@article{PolishTorus,
  author = {M. A. Abramowicz and M. Jaroszynski and M. Sikora},
  title = {{Relativistic, accreting disks}},
  year = {1978},
  journal = {Astron. Astrophys.},
  volume = {63},
  pages = {221},
  url = {https://articles.adsabs.harvard.edu/pdf/1978A%26A....63..221A}
}

@article{AbramowiczFragile,
  author = {M. A. Abramowicz and P. C. Fragile},
  title = {{Foundations of Black Hole Accretion Disk Theory}},
  year = {2013},
  journal = {Living Rev. Relativ.},
  volume = {16},
  pages = {1},
  eprint = {1104.5499},
  archivePrefix = {arXiv},
  primaryClass = {astro-ph.HE},
  note = {\href{https://arxiv.org/abs/1104.5499}{arXiv:1104.5499}},
  doi = {10.12942/lrr-2013-1}
}

@article{Younsi,
  author = {Z. Younsi and K. Wu and S. V. Fuerst},
  title = {{General relativistic radiative transfer: formulation and emission from structured tori around black holes}},
  year = {2012},
  journal = {Astron. Astrophys.},
  volume = {545},
  pages = {A13},
  eprint = {1207.4234},
  archivePrefix = {arXiv},
  primaryClass = {astro-ph.HE},
  note = {\href{https://arxiv.org/abs/1207.4234}{arXiv:1207.4234}},
  doi = {10.1051/0004-6361/201219599}
}

@article{KerrMOG,
  author = {Z. Zhang and S. Chen and J. Jing},
  title = {{Images of Kerr-MOG black holes surrounded by geometrically thick magnetized equilibrium tori}},
  year = {2024},
  journal = {JCAP},
  volume = {09},
  pages = {027},
  eprint = {2404.12223},
  archivePrefix = {arXiv},
  primaryClass = {gr-qc},
  note = {\href{https://arxiv.org/abs/2404.12223}{arXiv:2404.12223}}
}

@article{Komissarov,
  author = {S. S. Komissarov},
  title = {{Magnetized tori around Kerr black holes: analytic solutions with a toroidal magnetic field}},
  year = {2006},
  journal = {Mon. Not. R. Astron. Soc.},
  volume = {368},
  pages = {993--1000},
  eprint = {astro-ph/0601678},
  archivePrefix = {arXiv},
  note = {\href{https://arxiv.org/abs/astro-ph/0601678}{arXiv:astro-ph/0601678}},
  doi = {10.1111/j.1365-2966.2006.10183.x}
}

@article{VincentTorus,
  author = {F. H. Vincent and W. Yan and O. Straub and A. A. Zdziarski and M. A. Abramowicz},
  title = {{A magnetized torus for modeling Sgr A* millimeter images and spectra}},
  year = {2015},
  journal = {Astron. Astrophys.},
  volume = {574},
  pages = {A48},
  eprint = {1406.0353},
  archivePrefix = {arXiv},
  primaryClass = {astro-ph.GA},
  note = {\href{https://arxiv.org/abs/1406.0353}{arXiv:1406.0353}},
  doi = {10.1051/0004-6361/201424306}
}

@article{QMetricTorus,
  author = {G. D. Prada-M\'endez and F. D. Lora-Clavijo and J. M. Vel\'asquez-Cadavid},
  title = {{Synchrotron emitting Komissarov torus around naked singularities}},
  year = {2023},
  journal = {Class. Quantum Grav.},
  volume = {40},
  pages = {195011},
  eprint = {2308.09174},
  archivePrefix = {arXiv},
  primaryClass = {gr-qc},
  note = {\href{https://arxiv.org/abs/2308.09174}{arXiv:2308.09174}},
  doi = {10.1088/1361-6382/acf17e}
}

@article{PennaTorus,
  author = {R. F. Penna and A. Kulkarni and R. Narayan},
  title = {{A new equilibrium torus solution and GRMHD initial conditions}},
  year = {2013},
  journal = {Astron. Astrophys.},
  volume = {559},
  pages = {A116},
  eprint = {1309.3680},
  archivePrefix = {arXiv},
  primaryClass = {astro-ph.HE},
  note = {\href{https://arxiv.org/abs/1309.3680}{arXiv:1309.3680}},
  doi = {10.1051/0004-6361/201219666}
}

@article{DavisGammie,
  author = {S. W. Davis and C. F. Gammie},
  title = {{Covariant radiative transfer for black hole spacetimes}},
  year = {2020},
  journal = {Astrophys. J.},
  volume = {888},
  pages = {94},
  eprint = {1911.07950},
  archivePrefix = {arXiv},
  primaryClass = {astro-ph.HE},
  note = {\href{https://arxiv.org/abs/1911.07950}{arXiv:1911.07950}},
  doi = {10.3847/1538-4357/ab5950}
}

@article{Dexter,
  author = {J. Dexter},
  title = {{A public code for general relativistic, polarised radiative transfer around spinning black holes}},
  year = {2016},
  journal = {Mon. Not. R. Astron. Soc.},
  volume = {462},
  pages = {115--136},
  eprint = {1602.03184},
  archivePrefix = {arXiv},
  primaryClass = {astro-ph.HE},
  note = {\href{https://arxiv.org/abs/1602.03184}{arXiv:1602.03184}},
  doi = {10.1093/mnras/stw1526}
}

@article{IPole,
  author = {M. Mo\'scibrodzka and C. F. Gammie},
  title = {{ipole---semi-analytic scheme for relativistic polarized radiative transport}},
  year = {2018},
  journal = {Mon. Not. R. Astron. Soc.},
  volume = {475},
  pages = {43--54},
  eprint = {1712.03057},
  archivePrefix = {arXiv},
  primaryClass = {astro-ph.HE},
  note = {\href{https://arxiv.org/abs/1712.03057}{arXiv:1712.03057}},
  doi = {10.1093/mnras/stx3162}
}

@article{Gralla:2019xty,
  author = {S. E. Gralla and D. E. Holz and R. M. Wald},
  title = {{Black Hole Shadows, Photon Rings, and Lensing Rings}},
  year = {2019},
  journal = {Phys. Rev. D},
  volume = {100},
  pages = {024018},
  eprint = {1906.00873},
  archivePrefix = {arXiv},
  primaryClass = {astro-ph.HE},
  note = {\href{https://arxiv.org/abs/1906.00873}{arXiv:1906.00873}},
  doi = {10.1103/PhysRevD.100.024018}
}

@article{Perlick:2021aok,
  author = {V. Perlick and O. Y. Tsupko},
  title = {{Calculating black hole shadows: Review of analytical studies}},
  year = {2022},
  journal = {Phys. Rept.},
  volume = {947},
  pages = {1--39},
  eprint = {2105.07101},
  archivePrefix = {arXiv},
  primaryClass = {gr-qc},
  note = {\href{https://arxiv.org/abs/2105.07101}{arXiv:2105.07101}},
  doi = {10.1016/j.physrep.2021.10.004}
}

@article{TlemissovKovar,
  author = {A. Tlemissov and J. Kov\'a\v{r}},
  title = {{Egg-shaped sections of fluid structures around black holes}},
  year = {2024},
  journal = {Eur. Phys. J. C},
  volume = {84},
  pages = {1115},
  doi = {10.1140/epjc/s10052-024-13496-w}
}

@article{AlBadawi,
  author = {A. Al-Badawi and M. Halilsoy},
  title = {{Revisiting the rotating BR electromagnetic universe}},
  year = {2004},
  journal = {Nuovo Cim. B},
  volume = {119},
  pages = {931},
  url = {https://www.emu.edu.tr/stafffile/JOHXAGD.pdf}
}

@misc{GlobalKBR,
  author = {H. Ovcharenko},
  title = {{Global structure of Kerr--Newman--BR spacetime}},
  year = {2026},
  eprint = {2608.28265},
  archivePrefix = {arXiv},
  primaryClass = {gr-qc},
  note = {\href{https://arxiv.org/abs/2608.28265}{arXiv:2608.28265}}
}

@misc{ExtendedKNBR,
  author = {Y.-S. Zhou and W.-T. Fu and L.-M. Cao and R.-G. Cai},
  title = {{On the extended Kerr--Newman--Bertotti--Robinson spacetime: two black holes and a naked singularity in Bertotti--Robinson universe}},
  year = {2026},
  eprint = {2609.20626},
  archivePrefix = {arXiv},
  primaryClass = {gr-qc},
  note = {\href{https://arxiv.org/abs/2609.20626}{arXiv:2609.20626}}
}

@misc{KNBRThermo,
  author = {Z. Zhang and Z. Zhang and B. Chen},
  title = {{Thermodynamics of Kerr--Newman--Bertotti--Robinson black holes}},
  year = {2026},
  eprint = {2609.09816},
  archivePrefix = {arXiv},
  primaryClass = {gr-qc},
  note = {\href{https://arxiv.org/abs/2609.09816}{arXiv:2609.09816}}
}

@misc{NotesKBR,
  author = {Y.-S. Zhou and L.-B. Wu and M.-F. Ji and W.-T. Fu and L.-M. Cao and R.-G. Cai},
  title = {{Notes on Kerr--Bertotti--Robinson spacetime}},
  year = {2026},
  eprint = {2608.19136},
  archivePrefix = {arXiv},
  primaryClass = {gr-qc},
  note = {\href{https://arxiv.org/abs/2608.19136}{arXiv:2608.19136}}
}

@misc{WangISCO,
  author = {T. Wang},
  title = {{Innermost stable circular orbit of Kerr--Bertotti--Robinson black holes and inspirals from it: exact solutions}},
  year = {2025},
  eprint = {2508.04684},
  archivePrefix = {arXiv},
  primaryClass = {gr-qc},
  note = {\href{https://arxiv.org/abs/2508.04684}{arXiv:2508.04684}}
}

@article{Wald,
  author = {R. M. Wald},
  title = {{Black hole in a uniform magnetic field}},
  year = {1974},
  journal = {Phys. Rev. D},
  volume = {10},
  pages = {1680--1685},
  doi = {10.1103/PhysRevD.10.1680}
}

@article{Ernst,
  author = {F. J. Ernst},
  title = {{Black holes in a magnetic universe}},
  year = {1976},
  journal = {J. Math. Phys.},
  volume = {17},
  pages = {54--56},
  doi = {10.1063/1.522781}
}

@article{ErnstWild,
  author = {F. J. Ernst and W. J. Wild},
  title = {{Kerr black holes in a magnetic universe}},
  year = {1976},
  journal = {J. Math. Phys.},
  volume = {17},
  pages = {182},
  doi = {10.1063/1.522875}
}

@article{Stuchlik,
  author = {Z. Stuchl\'ik and S. Hled\'ik},
  title = {{Photon capture cones and embedding diagrams of the Ernst spacetime}},
  year = {1999},
  journal = {Class. Quantum Grav.},
  volume = {16},
  pages = {1377--1387},
  eprint = {0803.2536},
  archivePrefix = {arXiv},
  primaryClass = {gr-qc},
  note = {\href{https://arxiv.org/abs/0803.2536}{arXiv:0803.2536}}
}

@article{WangMelvin,
  author = {M. Wang and S. Chen and J. Jing},
  title = {{Kerr black hole shadows in Melvin magnetic field with stable photon orbits}},
  year = {2021},
  journal = {Phys. Rev. D},
  volume = {104},
  pages = {084021},
  eprint = {2104.12304},
  archivePrefix = {arXiv},
  primaryClass = {gr-qc},
  note = {\href{https://arxiv.org/abs/2104.12304}{arXiv:2104.12304}}
}

@article{JuniorMag,
  author = {H. C. D. L. Junior and P. V. P. Cunha and C. A. R. Herdeiro and L. C. B. Crispino},
  title = {{Shadows and lensing of black holes immersed in strong magnetic fields}},
  year = {2021},
  journal = {Phys. Rev. D},
  volume = {104},
  pages = {044018},
  eprint = {2104.09577},
  archivePrefix = {arXiv},
  primaryClass = {gr-qc},
  note = {\href{https://arxiv.org/abs/2104.09577}{arXiv:2104.09577}}
}

@misc{ZhuGuo,
  author = {H. Zhu and M. Guo},
  title = {{Polarized image of synchrotron radiation from hotspots in Schwarzschild--Melvin black hole spacetime}},
  year = {2022},
  eprint = {2205.04777},
  archivePrefix = {arXiv},
  primaryClass = {gr-qc},
  note = {\href{https://arxiv.org/abs/2205.04777}{arXiv:2205.04777}}
}

@article{HuQED,
  author = {Z. Hu and Z. Zhong and P.-C. Li and M. Guo and B. Chen},
  title = {{QED effect on a black hole shadow}},
  year = {2021},
  journal = {Phys. Rev. D},
  volume = {103},
  pages = {044057},
  eprint = {2012.07022},
  archivePrefix = {arXiv},
  primaryClass = {gr-qc},
  note = {\href{https://arxiv.org/abs/2012.07022}{arXiv:2012.07022}}
}

@article{ZhongQED,
  author = {Z. Zhong and Z. Hu and H. Yan and M. Guo and B. Chen},
  title = {{QED effects on Kerr black hole shadows immersed in uniform magnetic fields}},
  year = {2021},
  journal = {Phys. Rev. D},
  volume = {104},
  pages = {104028},
  eprint = {2108.06140},
  archivePrefix = {arXiv},
  primaryClass = {gr-qc},
  note = {\href{https://arxiv.org/abs/2108.06140}{arXiv:2108.06140}}
}

@article{EHTM87,
  author = {{Event Horizon Telescope Collaboration} and K. Akiyama and others},
  title = {{First M87 Event Horizon Telescope Results. I. The shadow of the supermassive black hole}},
  year = {2019},
  journal = {Astrophys. J. Lett.},
  volume = {875},
  pages = {L1},
  eprint = {1906.11238},
  archivePrefix = {arXiv},
  primaryClass = {astro-ph.GA},
  note = {\href{https://arxiv.org/abs/1906.11238}{arXiv:1906.11238}}
}

@article{EHTSgrA,
  author = {{Event Horizon Telescope Collaboration} and K. Akiyama and others},
  title = {{First Sagittarius A* Event Horizon Telescope Results. I. The shadow of the supermassive black hole in the center of the Milky Way}},
  year = {2022},
  journal = {Astrophys. J. Lett.},
  volume = {930},
  pages = {L12},
  eprint = {2311.08680},
  archivePrefix = {arXiv},
  primaryClass = {astro-ph.HE},
  note = {\href{https://arxiv.org/abs/2311.08680}{arXiv:2311.08680}}
}

@article{EHTPolarization,
  author = {{Event Horizon Telescope Collaboration} and K. Akiyama and others},
  title = {{First M87 Event Horizon Telescope Results. VII. Polarization of the ring}},
  year = {2021},
  journal = {Astrophys. J. Lett.},
  volume = {910},
  pages = {L12},
  eprint = {2105.01169},
  archivePrefix = {arXiv},
  primaryClass = {astro-ph.HE},
  note = {\href{https://arxiv.org/abs/2105.01169}{arXiv:2105.01169}}
}

@article{EHTMagnetic,
  author = {{Event Horizon Telescope Collaboration} and K. Akiyama and others},
  title = {{First M87 Event Horizon Telescope Results. VIII. Magnetic field structure near the event horizon}},
  year = {2021},
  journal = {Astrophys. J. Lett.},
  volume = {910},
  pages = {L13},
  eprint = {2105.01173},
  archivePrefix = {arXiv},
  primaryClass = {astro-ph.HE},
  note = {\href{https://arxiv.org/abs/2105.01173}{arXiv:2105.01173}}
}

@article{Carter,
  author = {B. Carter},
  title = {{Global structure of the Kerr family of gravitational fields}},
  year = {1968},
  journal = {Phys. Rev.},
  volume = {174},
  pages = {1559--1571},
  doi = {10.1103/PhysRev.174.1559}
}

@incollection{Bardeen,
  author = {J. M. Bardeen},
  title = {{Timelike and null geodesics in the Kerr metric}},
  year = {1973},
  booktitle = {Black Holes (Les Astres Occlus)},
  editor = {C. DeWitt and B. S. DeWitt},
  publisher = {Gordon and Breach},
  address = {New York},
  pages = {215--239},
  url = {https://ui.adsabs.harvard.edu/abs/1973blho.conf..215B/abstract}
}

@article{WalkerPenrose,
  author = {M. Walker and R. Penrose},
  title = {{On quadratic first integrals of the geodesic equations for type [22] spacetimes}},
  year = {1970},
  journal = {Commun. Math. Phys.},
  volume = {18},
  pages = {265--274},
  doi = {10.1007/BF01649445}
}

@article{Johannsen,
  author = {T. Johannsen},
  title = {{Photon rings around Kerr and Kerr-like black holes}},
  year = {2013},
  journal = {Astrophys. J.},
  volume = {777},
  pages = {170},
  eprint = {1501.02814},
  archivePrefix = {arXiv},
  primaryClass = {astro-ph.HE},
  note = {\href{https://arxiv.org/abs/1501.02814}{arXiv:1501.02814}}
}

@article{CunhaReview,
  author = {P. V. P. Cunha and C. A. R. Herdeiro},
  title = {{Shadows and strong gravitational lensing: a brief review}},
  year = {2018},
  journal = {Gen. Relativ. Gravit.},
  volume = {50},
  pages = {42},
  eprint = {1801.00860},
  archivePrefix = {arXiv},
  primaryClass = {gr-qc},
  note = {\href{https://arxiv.org/abs/1801.00860}{arXiv:1801.00860}}
}

@article{ChenReview,
  author = {S. Chen and J. Jing and W.-L. Qian and B. Wang},
  title = {{Black hole images: A review}},
  year = {2023},
  journal = {Sci. China Phys. Mech. Astron.},
  volume = {66},
  pages = {260401},
  eprint = {2301.00113},
  archivePrefix = {arXiv},
  primaryClass = {astro-ph.HE},
  note = {\href{https://arxiv.org/abs/2301.00113}{arXiv:2301.00113}}
}

@article{GrallaLensing,
  author = {S. E. Gralla and A. Lupsasca},
  title = {{Lensing by Kerr black holes}},
  year = {2020},
  journal = {Phys. Rev. D},
  volume = {101},
  pages = {044031},
  eprint = {1910.12873},
  archivePrefix = {arXiv},
  primaryClass = {gr-qc},
  note = {\href{https://arxiv.org/abs/1910.12873}{arXiv:1910.12873}}
}

@article{HouKN,
  author = {Y. Hou and P. Liu and M. Guo and H. Yan and B. Chen},
  title = {{Multi-level images around Kerr--Newman black holes}},
  year = {2022},
  journal = {Class. Quantum Grav.},
  volume = {39},
  pages = {194001},
  eprint = {2203.02755},
  archivePrefix = {arXiv},
  primaryClass = {gr-qc},
  note = {\href{https://arxiv.org/abs/2203.02755}{arXiv:2203.02755}},
  doi = {10.1088/1361-6382/ac8860}
}

@article{HouKerrMelvin,
  author = {Y. Hou and Z. Zhang and H. Yan and M. Guo and B. Chen},
  title = {{Image of a Kerr--Melvin black hole with a thin accretion disk}},
  year = {2022},
  journal = {Phys. Rev. D},
  volume = {106},
  pages = {064058},
  eprint = {2206.13744},
  archivePrefix = {arXiv},
  primaryClass = {gr-qc},
  note = {\href{https://arxiv.org/abs/2206.13744}{arXiv:2206.13744}},
  doi = {10.1103/PhysRevD.106.064058}
}

@article{ChaelInner,
  author = {A. Chael and M. D. Johnson and A. Lupsasca},
  title = {{Observing the inner shadow of a black hole: A direct view of the event horizon}},
  year = {2021},
  journal = {Astrophys. J.},
  volume = {918},
  pages = {6},
  eprint = {2106.00683},
  archivePrefix = {arXiv},
  primaryClass = {astro-ph.HE},
  note = {\href{https://arxiv.org/abs/2106.00683}{arXiv:2106.00683}}
}

@article{GatesBlueshift,
  author = {D. E. A. Gates and S. Hadar and A. Lupsasca},
  title = {{Maximum observable blueshift from circular equatorial Kerr orbiters}},
  year = {2020},
  journal = {Phys. Rev. D},
  volume = {102},
  pages = {104041},
  eprint = {2009.03310},
  archivePrefix = {arXiv},
  primaryClass = {gr-qc},
  note = {\href{https://arxiv.org/abs/2009.03310}{arXiv:2009.03310}}
}

@article{GellesPolar,
  author = {Z. Gelles and E. Himwich and D. C. M. Palumbo and M. D. Johnson},
  title = {{Polarized image of equatorial emission in the Kerr geometry}},
  year = {2021},
  journal = {Phys. Rev. D},
  volume = {104},
  pages = {044060},
  eprint = {2105.09440},
  archivePrefix = {arXiv},
  primaryClass = {gr-qc},
  note = {\href{https://arxiv.org/abs/2105.09440}{arXiv:2105.09440}}
}

@article{Palumbo,
  author = {D. C. M. Palumbo and G. N. Wong and B. S. Prather},
  title = {{Discriminating accretion states via rotational symmetry in simulated polarimetric images of M87}},
  year = {2020},
  journal = {Astrophys. J.},
  volume = {894},
  pages = {156},
  eprint = {2004.01751},
  archivePrefix = {arXiv},
  primaryClass = {astro-ph.HE},
  note = {\href{https://arxiv.org/abs/2004.01751}{arXiv:2004.01751}}
}

@article{ChangObserver,
  author = {Z. Chang and Q.-H. Zhu},
  title = {{The observer-dependent shadow of the Kerr black hole}},
  year = {2021},
  journal = {JCAP},
  volume = {09},
  pages = {003},
  eprint = {2104.14221},
  archivePrefix = {arXiv},
  primaryClass = {gr-qc},
  note = {\href{https://arxiv.org/abs/2104.14221}{arXiv:2104.14221}}
}

@article{Lindquist,
  author = {R. W. Lindquist},
  title = {{Relativistic transport theory}},
  year = {1966},
  journal = {Ann. Phys.},
  volume = {37},
  pages = {487--518},
  doi = {10.1016/0003-4916(66)90207-7}
}

@article{HouPolar,
  author = {Y. Hou and J. Huang and Minyong Guo and Y. Mizuno and B. Chen},
  title = {{Near-horizon polarization as a diagnostic of black hole spacetime}},
  year = {2025},
  journal = {Astrophys. J. Lett.},
  volume = {988},
  pages = {L51},
  eprint = {2409.07248},
  archivePrefix = {arXiv},
  primaryClass = {gr-qc},
  doi = {10.3847/2041-8213/adee09},
  note = {\href{https://arxiv.org/abs/2409.07248}{arXiv:2409.07248}}
}

@article{HouMagnetofluids,
  author = {Y. Hou and Z. Zhang and M. Guo and B. Chen},
  title = {{A new analytical model of magnetofluids surrounding rotating black holes}},
  year = {2024},
  journal = {JCAP},
  volume = {02},
  pages = {030},
  eprint = {2309.13304},
  archivePrefix = {arXiv},
  primaryClass = {gr-qc},
  note = {\href{https://arxiv.org/abs/2309.13304}{arXiv:2309.13304}}
}

@article{ZhangThick,
  author = {Z. Zhang and Y. Hou and M. Guo and B. Chen},
  title = {{Imaging thick accretion disks and jets surrounding black holes}},
  year = {2024},
  journal = {JCAP},
  volume = {05},
  pages = {032},
  eprint = {2401.14794},
  archivePrefix = {arXiv},
  primaryClass = {astro-ph.HE},
  note = {\href{https://arxiv.org/abs/2401.14794}{arXiv:2401.14794}}
}

@article{Cassing,
  author = {M. Cassing and L. Rezzolla},
  title = {{Equilibrium non-selfgravitating tori around black holes in parameterised spherically symmetric spacetimes}},
  year = {2023},
  journal = {Mon. Not. R. Astron. Soc.},
  volume = {522},
  pages = {2415},
  eprint = {2302.09135},
  archivePrefix = {arXiv},
  primaryClass = {gr-qc},
  note = {\href{https://arxiv.org/abs/2302.09135}{arXiv:2302.09135}}
}

@misc{MaLuSeeds,
  author = {Liang Ma and H. L{\"u}},
  title = {{Ricci-flat Seeds, Trombone Rescaling and Magnetizing Kerr Black Holes}},
  year = {2026},
  eprint = {2610.30256},
  archivePrefix = {arXiv},
  primaryClass = {gr-qc},
  note = {\href{https://arxiv.org/abs/2610.30256}{arXiv:2610.30256}}
}

@article{KozlowskiTorus,
  author = {M. Kozlowski and M. Jaroszynski and M. A. Abramowicz},
  title = {{The analytic theory of fluid disks orbiting the Kerr black hole}},
  journal = {Astron. Astrophys.},
  volume = {63},
  pages = {209--220},
  year = {1978},
  url = {https://articles.adsabs.harvard.edu/pdf/1978A%26A....63..209K}
}

@article{StraubIonTorus,
  author = {O. Straub and F. H. Vincent and M. A. Abramowicz and E. Gourgoulhon and T. Paumard},
  title = {{Modelling the black hole silhouette in Sagittarius A* with ion tori}},
  journal = {Astron. Astrophys.},
  volume = {543},
  pages = {A83},
  year = {2012},
  doi = {10.1051/0004-6361/201219209},
  eprint = {1203.2618},
  archivePrefix = {arXiv},
  primaryClass = {astro-ph.HE},
  note = {\href{https://arxiv.org/abs/1203.2618}{arXiv:1203.2618}}
}

@article{SuKerrEmission,
  author = {Y. Su and Minyong Guo and H. Yan and B. Chen},
  title = {{Photon emissions from Kerr equatorial geodesic orbits}},
  journal = {Eur. Phys. J. C},
  volume = {84},
  pages = {733},
  year = {2024},
  doi = {10.1140/epjc/s10052-024-13113-w}
}

@misc{Coport,
  author = {J. Huang and L. Zheng and Minyong Guo and B. Chen},
  title = {{Coport: A New Public Code for Polarized Radiative Transfer in a Covariant Framework}},
  year = {2024},
  eprint = {2407.10431},
  archivePrefix = {arXiv},
  primaryClass = {astro-ph.HE},
  doi = {10.48550/arXiv.2407.10431},
  note = {\href{https://arxiv.org/abs/2407.10431}{arXiv:2407.10431}}
}
\endgroup

\end{document}